\documentclass[onecolumn,a4paper,12pt,sort&compress]{elsarticle}

\usepackage{ifpdf}
\usepackage[utf8]{inputenc}
\usepackage{hyperref}
\usepackage[euler]{textgreek}
\usepackage[draft]{minted}

\usepackage{tabto}
\usepackage{multicol}
\usepackage{graphicx}
\usepackage{float}
\usepackage{graphics}
\usepackage{pdflscape}
\usepackage{multirow}
\usepackage{makecell}
\usepackage{adjustbox}
\usepackage{rotating}
\usepackage{paralist}  
\usepackage{longtable}  
\usepackage{comment}  
\usepackage{colortbl}
\usepackage{enumitem}

\usepackage{xparse}

\usepackage{listings,xcolor}
\usepackage{multicol}
\usepackage{graphicx}
\usepackage{float}
\usepackage{graphics}
\usepackage{pdflscape}
\usepackage{multirow}
\usepackage{makecell}
\usepackage{adjustbox}
\usepackage{rotating}
\usepackage{paralist} 
\usepackage{longtable} 
\usepackage{comment} 
\usepackage{colortbl}
\usepackage{tikz}
\usepackage{hyperref} 
\usepackage{tcolorbox}
\usepackage{xparse}

\usepackage[euler]{textgreek}
\definecolor{iscte-iul-palette}{HTML}{14BFB8}
\usepackage{xspace}
\usepackage[linesnumbered,ruled,vlined]{algorithm2e}
\usepackage{amssymb}
\usepackage{enumitem}
\usepackage[percent]{overpic}
\usepackage{tablefootnote}
\usepackage[euler]{textgreek}

\usepackage{tcolorbox}
\usepackage[position=above, skip=0pt]{caption}

\journal{Journal of Systems and Software}

\usepackage{numcompress}
\NewDocumentCommand{\plot}{m O{} O{12} O{H} O{trim=0cm 0cm 0cm 0cm} }{
\begin{figure}[#4]
\centering
\includegraphics[width=#3cm,clip,#5]{#1}
\caption{#2}
\label{#1}
\end{figure}
}

\usepackage{tikz}
\usetikzlibrary{svg.path}
\definecolor{orcidlogocol}{HTML}{A6CE39}
\tikzset{
	orcidlogo/.pic={
		\fill[orcidlogocol] svg{M 100 100 C 100 72.4 77.6 50 50 50 C 22.3 50 0 72.4 0 100 C 0 127.6 22.3 150 50 150 C 77.6 150 100 127.6 100 100 Z};
		\fill[white] svg{M 33.666 77.27 L 27.651 77.27 L 27.651 119.103 L 33.666 119.103 L 33.666 100.198 L 33.666 77.27 Z};
		\fill[white] svg{M 42.493 119.103 L 58.742 119.103 C 74.21 119.103 81.007 108.049 81.007 98.167 C 81.007 87.425 72.609 77.23 58.821 77.23 L 42.493 77.23 L 42.493 119.103 Z M 48.509 82.66 L 58.078 82.66 C 71.71 82.66 74.835 93.011 74.835 98.167 C 74.835 106.565 69.484 113.673 57.766 113.673 L 48.509 113.673 L 48.509 82.66 Z};
		\fill[white] svg{M 34.603 127.813 C 34.603 125.665 32.846 123.868 30.658 123.868 C 28.471 123.868 26.713 125.665 26.713 127.813 C 26.713 130.001 28.471 131.758 30.658 131.758 C 32.846 131.758 34.603 129.961 34.603 127.813 Z};
	}
}
\def\orcid#1{%
	\href{https://orcid.org/#1}{%
		\begin{tikzpicture}[baseline]
			\pic[scale=0.1,yshift=-7em] {orcidlogo};
		\end{tikzpicture}
	}%
}

\begin{document}

\begin{frontmatter}

\title{Profiling Software Developers with Process Mining and N-Gram Language Models}


\author[ISTAR]{João Caldeira
\textsuperscript{\orcid{0000-0003-0960-0179}}}
\ead{jcppc@iscte-iul.pt}


\author[ISTAR,DCTI]{Fernando Brito e Abreu
\textsuperscript{\orcid{0000-0002-9086-4122}}}
\ead{fba@iscte-iul.pt}

\author[CISUC]{Jorge Cardoso \textsuperscript{\orcid{0000-0001-8992-3466}}}
\ead{jcardoso@dei.uc.pt}

\author[DCTI,INESC]{Ricardo Ribeiro
\textsuperscript{\orcid{0000-0002-2058-693X}}}
\ead{ricardo.ribeiro@iscte-iul.pt}

\author[COPPE,ISTAR]{Claudia Werner
\textsuperscript{\orcid{0000-0002-4231-9621}}}
\ead{werner@cos.ufrj.br}

\address[ISTAR]{ISTAR, Iscte - Instituto Universitário de Lisboa, Portugal}
\address[DCTI]{DCTI, Iscte - Instituto Universitário de Lisboa, Portugal}
\address[CISUC]{CISUC, University of Coimbra, Portugal and Huawei Munich Research Center, Germany}
\address[COPPE]{COPPE, Federal University of Rio de Janeiro, Brazil}
\address[INESC]{INESC-ID Lisboa, Portugal}

\begin{abstract}

\noindent\textit{\textbf{Context}}:
Profiling developers is challenging since many factors, such as their skills, experience, development environment and behaviors, may influence a detailed analysis and the delivery of coherent interpretations.

\noindent\textit{\textbf{Objective}}: 
We aim at profiling software developers by mining their software development process. To do so, we performed a controlled experiment where, in the realm of a Python programming contest, a group of developers had the same well-defined set of requirements specifications and a well-defined sprint schedule. Events were collected from the PyCharm IDE, and from the Mooshak automatic jury where subjects checked-in their code. 

\noindent\textit{\textbf{Method}}: 
We used n-gram language models and text mining to characterize developers' profiles, and process mining algorithms to discover their overall workflows and extract the correspondent metrics for further evaluation.

\noindent\textit{\textbf{Results}}: 
Findings show that we can clearly characterize with a coherent rationale most developers, and distinguish the top performers from the ones with more challenging behaviors. This approach may lead ultimately to the creation of a catalog of software development process smells.

\noindent\textit{\textbf{Conclusions}}: 
The profile of a developer provides a software project manager a clue for the selection of appropriate tasks he/she should be assigned. With the increasing usage of low and no-code platforms, where coding is automatically generated from an upper abstraction layer, mining developer's actions in the development platforms is a promising approach to early detect not only behaviors but also assess project complexity and model effort.

\end{abstract}



\end{frontmatter}



\section{Introduction}
\label{sec:introduction}

\subsection{Motivation}

Software development can be characterized as a socio-technical phe\-nom\-e\-non \cite{Fuggetta2014SoftwareProcess}. Understanding the actual dependencies between development tasks and teams’ behaviors to fulfill them is a serious challenge for most software project managers concerned with the allocation and coordination of resources \cite{Herbsleb2016BuildingAward}. Being able to group developers with similar behaviors, for instance, based on the time they spent on each activity or working on a specific artifact, is a step forward in that understanding. This requires analyzing developers' traces (i.e. executed actions/commands) within the IDE.

Traditional process mining techniques come to the rescue of such concerns. However, within the software development context, the latter usually assumes a structured and noise-free input and produces spaghetti-like processes. As such, a lot of variances may mislead the results and correspondent interpretation \cite{Hompes2015DiscoveringClustering}. Process variant analysis is a research stream within the process mining domain. In the last decade, several novel approaches to effectively mine process variants have been proposed \cite{Bolt2017FindingPaper, Bolt2018, Taymouri2020BusinessLogs}. The latter evolved to detect the existence of similarities and differences in behaviors within a common business process, which can be considered as ``fingerprints'' left by process instances \cite{VanDerAalst2004, Cook1996}.

Applying process mining algorithms on large event logs, containing a significant number of cases and events, usually requires the use of powerful computational systems and, even then, may lead to long processing times. Process variant comparison techniques, in particular due to massive manipulation of vectors and matrices, are computationally heavy. Software development event logs, generated from IDE usage, often have events at the thousands, hundreds of different activities, hierarchical states, and many different resources associated with events. Therefore, the aforementioned performance problem is usually noticeable. Natural language techniques can mitigate it by performing initial filtering, aggregation of events and in finding local regularities. Even if event aggregation is not desired in mining processes, the trade-off between the practical aspects versus the accuracy of certain algorithms should be carefully evaluated in the software development realm. 

In this paper, we propose an approach to profile developers using a stack of text mining to express developers' fingerprints, and process mining to discover, model and assist in hypothesis evaluation regarding their workflows. We used events collected from the IDE during development sessions as input for the unsupervised learning techniques and process mining algorithms. Additionally, since the process of coding can be represented as a grammar with a specific semantic \cite{Hindle2012OnSoftware}, we find it useful to assess how similar this grammar is to a natural language, and in finding optimal parameters for the text mining algorithms.

A development session executed by one developer at his/her IDE can be considered an instance of a process, where the goal is to produce a software product or maintain an existing one. Its workflow of activities depends on many factors, such as the development methodology, program design or individual experience \cite{Caldeira2019AssessingMining}. Furthermore, developers are usually free to produce code without a referential model or guidelines on how to execute the coding tasks and, most often without any intelligent guidance from traditional development tools. This poses challenges when one wants to detect similarly or deviating programming profiles to assess productivity and optimize resource planning. 

To validate our approach for profiling developers, while controlling for spurious effects, we performed a controlled experiment where, in the realm of a \href{https://sites.google.com/iscte-iul.pt/pythacon}{Python programming contest}, a group of developers had the same well-defined set of requirements specifications and a well-defined sprint schedule. Events were collected from the \href{https://www.jetbrains.com/pycharm/}{PyCharm IDE}, and from the \href{https://mooshak.dcc.fc.up.pt/}{Mooshak automatic judge} where subjects checked-in their code stepwise.

\subsection{Using students as surrogates for professional software developers}

In this study we use students as surrogates for professional software developers. Therefore it is worth reviewing the discussion in the literature on using students as surrogates for professionals. 

Almost half a century ago, the practice of using students in research was already widespread, due to the convenience of their availability and usual willingness to participate in experiments. For instance, in consumer behavior (marketing) studies, researchers tested whether students could be used as consumer surrogates, but results were inconclusive \cite{Enis1972Students, Shuptrine1975Validity, Hampton1979Students}. Also since the seventies, as reported in \cite{Remus1989Using}, students have been used as surrogates for managers on decision support systems (DSS). The same study reports that undergraduate students were more used than graduate students, which could be a validity hindrance in that case, since graduate students are more closer to managers in age, maturity and education. 

Students have also been extensively used as surrogates in Software Engineering studies. For instance, a study carried out with students on detection methods for software requirements inspections \cite{Porter1995Comparing} was replicated with similar results using professionals as subjects \cite{Porter1998Comparing}. Another study on lead-time impact assessment for software development projects did not find significant differences between students and professionals \cite{Host2000Using}. In \cite{Runeson2003Using}, the performance in Personal Software Process (PSP) improvement tasks was compared between freshmen students, graduate students, and industry people, and again no significant differences were found between the three groups. Two separate studies in Requirements Engineering provided somehow complementary conclusions. While in \cite{Berander2004Using}  definitive conclusions about the suitability of students in projects could not be drawn, in \cite{Svahnberg2008Using} the authors argue that it may be possible to influence students to provide answers that are in line with industrial practice, although it was not clear under which conditions could that influence be exerted in empirical investigations. A systematic literature review on using students as surrogates for professionals can be found in \cite{Kotakonda2012Are}. The author concludes that many factors influence the results of experimental studies such as the number of subjects, nature of tasks and previous experience on that, motivation levels of subjects, training provided, and incentives for participation in the experiment. In other words, the appropriateness of students as surrogates for professionals depends on current study conditions. In section \ref{sec:ExternalValidity5}, we argue why this may hold in our study.

\subsection{Contributions}

The main objectives for this work are the following:
\begin{itemize}

\item To evaluate if software development sessions can be mined as any natural language;

\item To assess if coherent development fingerprints can be discovered from an event log containing developer's IDE interactions and submission of answers to several coding problems;

\item To appraise the impact of individual behaviors in the outcome of a programming task given a group of developers.




\end{itemize}

The remainder of this paper is organized as follows: section \ref{sec:Background5} provides background related to the research area and emphasizes the need for the proposed approach. Subsequent sections, outline the related work in section \ref{sec:RelatedWork5}, detail the methodology and experiment setup in section \ref{sec:study-setup5} and present the results, its corresponding analysis and implications in section \ref{sec:study-results5}. Threats to validity are presented in section \ref{sec:Threatsvalidity5} and the concluding comments and future work in section \ref{sec:Conclusions5}.

\section{Background}
\label{sec:Background5}

\subsection{Language models}
\label{language-models}
Natural languages (e.g., English, Portuguese, etc.) possess a rich vocabulary and therefore are complex and powerful. A programming language or a sequence of development actions in plain English, as seen in Figure \ref{chapter7-1.pdf}\footnote{This word cloud, where the size of each word is proportional to its relative frequency, was generated from data collected during the validation experiment of our proposed approach.}, is an artificial language but is expected to follow the same principles of a natural language. The rationale is that although a given piece of software is written with an artificial language, it is a natural product of the human mind as prose or poetry in natural language \cite{Hindle2012OnSoftware}. As such, it is also amenable to statistical analysis like the ones performed in the area known as ``text mining'', where natural language processing (NLP) algorithms and analytical methods are used.

We argue that development sessions viewed as a sequence of actions like those in Figure \ref{chapter7-1.pdf} and represented by a well-defined vocabulary, can be regarded from the same perspective. 
In this paper, we describe a novel method to detect different developers' profiles based on models built from development interactions using n-gram probabilistic language models \cite{Jurafsky2020Speech}. Furthermore, we combine these unsupervised learning models, which present a good fit in capturing local regularities in text data, and process mining algorithms, which are known to perform well in the modelling of complex business processes.



\plot{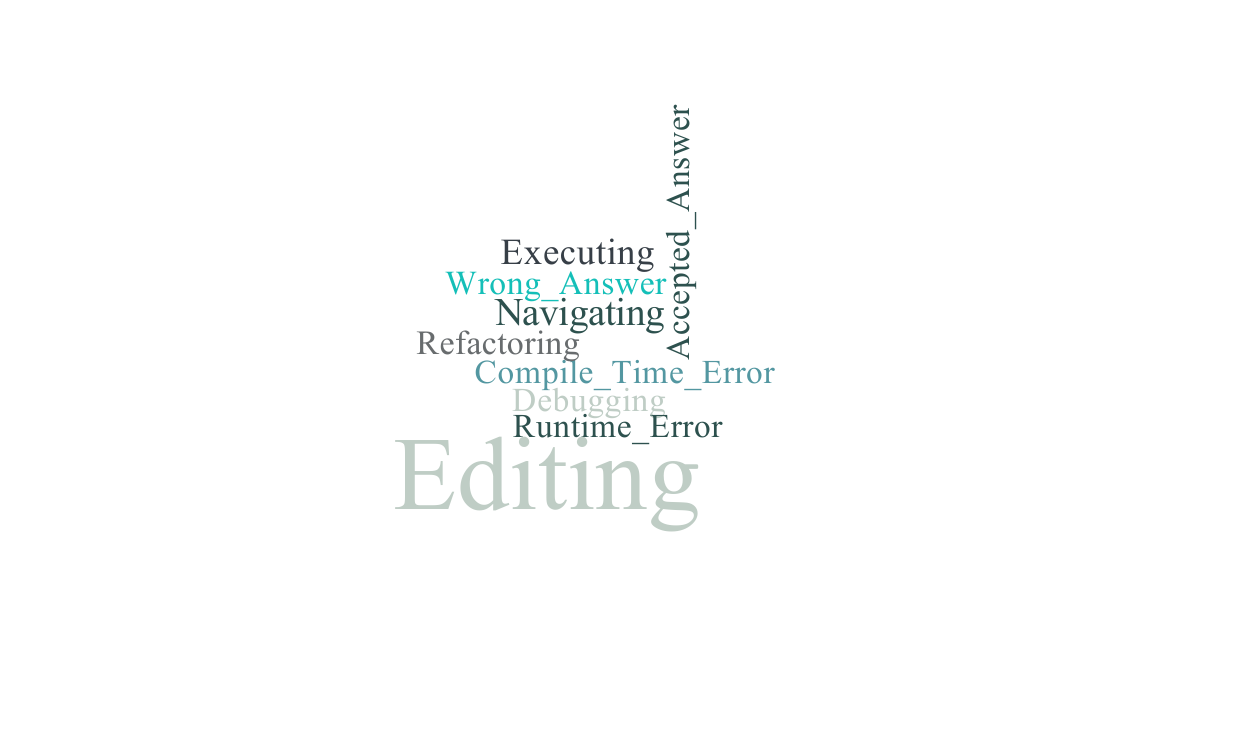}[Word cloud with example of frequent activities on interactions in PyCharm and submissions to Mooshak][5][ht!][trim=4cm 2cm 4.5cm 1cm]

\subsection{Topic modeling}
\label{topic-modeling}

Understanding unstructured data is a major challenge in software development, and having a predefined data model is not a common scenario when dealing with such type of data. Moreover, those data are typically text-heavy. As such, topic modeling has become one of the most used methods to mine software repositories \cite{Chen2011ALines}.

Topic modeling is a method for unsupervised classification of documents, by modeling each document as a mixture of topics and each topic as a mixture of words \cite{Nguyen2012DuplicateModeling}. Despite some limitations, such as the order of the documents, it is frequently used to build models from unstructured textual data, as it presents an effective means of data mining where subjects represent documents or even a textual representation of actions executed in certain contexts \cite{Agrawal2018WhatEngineering}.

Within the most prevalent methods used to mine software repositories, we find algorithms such as LDA (Latent Dirichlet Allocation) and many of its variations, LSI (Latent Semantic Indexing), LSA (Latent Semantic Analysis), PLSI (Probabilistic Latent Semantic Indexing), and ICA (Independent Component Analysis) \cite{Chen2016ARepositories, Blei2003LatentAllocation}. These algorithms are used to cluster documents, identify features, derive source code metrics for bugs prediction, assess code evolution, trace links between pairs of artifacts and detect code clones, among other things \cite{Menzies2015TheMethods, Chen2011ALines}.

\subsection{Software development process mining}
\label{process-mining}

Process modeling is a persistent topic in the research literature concerned with software development practices. The analysis of fingerprints in event logs \cite{Taymouri2020BusinessLogs}, the discovery of deviating cases using trace clustering \cite{Hompes2015DiscoveringClustering} and mining of sequences of developers interactions \cite{Damevski2017MiningSmells} are examples of topics covered by researchers to overcome or mitigate recurrent problems. However, often the suggested solutions are complex and difficult to automate in a coherent software development process mining pipeline. These constraints led researchers to highlight that software analytics does not need to be hard and, on the contrary, it can and should be simplified \cite{Fu2017EasyLearning, Agrawal2019HowAnalytics, Fu2016TuningNecessary}.

In Table \ref{table:sdpm-topic-modelling}, we present a comparison of typical text mining characteristics and purposes, along with how we view topic modeling applied in software development process mining.

Software developers execute a stream of actions/commands when using their IDE. Those commands, seen as a work session, can also be represented textually as a narrative along the timeline. We expect that stream to contain the semantics required to identify different developer profiles. Therefore, logs containing a sequence of IDE commands/actions can be mined with topic modeling as any other document would be in searching for different topics. In this context, we are searching for different behaviors such as programming styles or patterns of IDE usage.

\begin{table}[H]
\scriptsize
\caption{Traditional Text Mining (TM)  vs. Software Development Process Mining (SDPM)}
\label{table:sdpm-topic-modelling}
\centering
\begin{tabular}{lll}
	\hline\noalign{\smallskip}
			\textbf{} & \textbf{TM} & \textbf{SDPM}\\
	\noalign{\smallskip}\hline\noalign{\smallskip}
	
	\multicolumn{3}{l}{\cellcolor{gray!10}\textbf{Inputs}}\\[0.1cm]
	\textbf{Corpus} & Documents/Articles & Development Work Sessions\\[0.1cm]
	\textbf{Document} & Mixture of Topics & Mixture of Behaviours\\[0.1cm]
	\textbf{Topic} & Frequent Words/Terms & Frequent Actions/Commands\\[0.2cm]
    \multicolumn{3}{l}{\cellcolor{gray!10}\textbf{Outputs}}\\[0.1cm]
	\textbf{Discovers} & Distinct Subjects/Topics & Development Patterns\\[0.1cm]
	\textbf{Usefulness} & Identify Social Trends & Optimize Resource Allocation\\
	\textbf{} & Frame Research Interests & Detect Practices Deviations\\
	\textbf{} & Sentiment Analysis & Forensic Project Analysis\\[0.2cm]
   \noalign{\smallskip}\hline
\end{tabular}
\end{table}

\subsubsection{Preliminary definitions}
\label{sec:preliminaries}

To justify the usefulness of collecting IDE events, and provide context to our proposal, we introduce in this section some preliminary definitions required to understand concepts such as development actions, development sessions, development actions repository and development profiles.\\

\noindent\textbf{Definition 1. Development action} 

\begin{itemize}[leftmargin=*,noitemsep,topsep=0pt]
    \item \textit{A development action is an event defined as a tuple (a, c, t, (p\textsubscript{1},v\textsubscript{1}),...,(p\textsubscript{n},v\textsubscript{n})) where a is the command action or process activity name, c is a development session or case id, t is the timestamp and the set (p\textsubscript{1},v\textsubscript{1}),...,(p\textsubscript{n},v\textsubscript{n}) (where n $\geqslant0$) contains the event or case properties/attributes and corresponding values, such as developer location, operating system or IDE type.}
    
    \item \textit{A development action is defined by a token t included in the development session vocabulary to be formed by a set containing all the possible IDE commands, denoted by V:
    \subitem V = (t\textsubscript{0},t\textsubscript{1},...t\textsubscript{n}) : $\forall$t $\in$ V, t = $\langle$ide\_command\_or\_activity$\rangle$}\\
\end{itemize}

\noindent\textbf{Definition 2. Development session} 
\begin{itemize}[leftmargin=*,noitemsep,topsep=0pt]
   
    \item \textit{A development session is a trace, defined by a non-empty sequence $\sigma$=e\textsubscript{1},...,e\textsubscript{n} of command actions such that $\forall$i,j $\in$ [1..n] e\textsubscript{i}.c=e\textsubscript{j}.c.}

    \item \textit{A development session is defined by a sentence formed by a set of tokens from vocabulary V, denoted by:
    \subitem $\omega$ = (t\textsubscript{0},t\textsubscript{1},...t\textsubscript{n}) : $\forall$t $\in$ $\omega$, t $\in$ V}\\
 \end{itemize}

\noindent\textbf{Definition 3. Development actions repository} 
\begin{itemize}[leftmargin=*,noitemsep,topsep=0pt]
    \item \textit{A repository of actions or event log is a set of development actions mapped to a variable number of development sessions, defined as L=$\sigma$\textsubscript{1},...,$\sigma$\textsubscript{n}}.
    \item \textit{We consider an event log a set of sequential tokens t from the vocabulary V, where t can be repeated.}\\
\end{itemize}

\noindent\textbf{Definition 4. Development profile} 
\begin{itemize}[leftmargin=*,noitemsep,topsep=0pt]
   \item \textit{An event log L can be partitioned into a finite set of groups called process variants or, in our case, profiles or fingerprints, $\varsigma$\textsubscript{1},$\varsigma$\textsubscript{2},...,$\varsigma$\textsubscript{n}, where $\exists$p such that $\forall$ $\varsigma$\textsubscript{k} and $\forall$ $\sigma$\textsubscript{i},$\sigma$\textsubscript{j} $\in$ $\varsigma$\textsubscript{k}, $\sigma$\textsubscript{i}.p=$\sigma$\textsubscript{j}.p.}\\
\end{itemize}

\indent The definition of process variant emphasizes that process executions in the same group must have the same value for a given attribute, and each process execution belongs uniquely to a process variant. In our approach, the same value for a given attribute will be dynamically computed and concatenated into the original dataset. The algorithms to model processes will then be based on this clustering action.\\

\noindent \textbf{Definition 5. N-gram language models}. 
\begin{itemize}[leftmargin=*,noitemsep,topsep=0pt]
\item \textit{A language model is a statistical model that allows computing the probability of a sentence, or predict the next word in a sentence for a given language \cite{Brown1992Class-BasedLanguage}. From a generative perspective, all sentences of a (natural) language can be described in terms of the product of a set of conditional probabilities \cite{Santos2017StepwiseModels}. Hence, the probability of a sentence $\omega$ = (t\textsubscript{0},t\textsubscript{1},...t\textsubscript{n}) is given by :}
\subitem \textit{ P($\omega$) = P(t\textsubscript{0})P(t\textsubscript{1}$|$t\textsubscript{0})P(t\textsubscript{2}$|$t\textsubscript{0}t\textsubscript{1})...P(t\textsubscript{n}$|$t\textsubscript{0}t\textsubscript{1}...t\textsubscript{n-1}) }\\
\end{itemize}





\section{Related Work}
\label{sec:RelatedWork5}

\subsection{Natural language models}

\subsubsection{Language modeling}

The use of natural language models was presented as an approach to recommend analogical libraries based on a knowledge base of analogical libraries mined from tags of millions of Stack Overflow questions \cite{Chen2019WhatsDiscussions}. This approach used a combination of a word embedding technique and domain-specific relational and categorical knowledge mined from Stack Overflow. Evidence showed that accurate recommendation of analogical libraries is not only possible but also a desirable solution. 

A system that assists developers in API usage with code completion recommendation, using a n-gram probabilistic language model, supported by API sentences extracted from source code corpora, is described in \cite{Santos2017StepwiseModels}.

\subsubsection{Topic modeling}

A survey on the use of topic models when mining software repositories is presented in \cite{Chen2016ARepositories}. The authors found that only a limited number of software engineering tasks were being targeted, and researchers use topic models as black boxes without fully evaluating their fundamental assumptions. Finally, they provide guidelines on how to apply topic models to specific software engineering tasks.

With the goal of predicting future developer behavior in the IDE and to make better recommendations to developers, \cite{Damevski2018PredictingModels} used topic models and specifically applied the Temporal Latent Dirichlet Allocation algorithm on two large interaction datasets for two different IDEs, Microsoft Visual Studio and ABB Robot Studio. The authors concluded that the approach was promising for both predicting future IDE commands and producing empirically-interpretable observation.

An approach to detect duplicate bug reports, using information retrieval and topic modeling, namely LDA, was presented in \cite{Nguyen2012DuplicateModeling}. The latter revealed an improvement of up to 20\% in accuracy, when compared to other state-of-the-art approaches.

A study of software logging using topic models, with the aim of understanding the relationship between the topics of a code snippet and the likelihood of a code snippet being logged (i.e. to contain a logging statement) is described in \cite{Li2018StudyingModels}. The findings highlight the topics containing valuable information that can help to guide and driving developers' logging decisions. A similar approach is presented in \cite{Ye2017TheOverflow}, based on the structure and dynamics of knowledge network in domain-specific Q\&A sites, particularly on Stack Overflow.

A large-scale study on security-related questions on Stack Overflow was presented in \cite{Yang2016WhatPosts}. Two heuristics were used to extract the questions that are related to security from the dataset based on the posts' tags. Later, to cluster different security-related questions based on their texts, the authors used LDA tuned with a Genetic Algorithm (GA).

\subsection{Mining software repositories}

An application of mining three software repositories: team wiki (used during requirement engineering), version control system (development and maintenance), and issue tracking system (corrective and adaptive maintenance) in the context of an undergraduate Software Engineering course was presented in \cite{Mittal2014ProcessCourse}. Visualizations and metrics provided insights into practices and procedures followed during various phases of a software development life-cycle, granting a multi-faceted view to the instructor and serving as a feedback tool on the development process and quality by students. Examples of insights produced by mining software repositories include understanding and assessing: (i) the degree of individual contributions in a team, (ii) the quality of commit messages, (iii) the intensity and consistency of commit activities, (iv) the trend and quality of the bug fixing process, (v) the component and developer entropy and, (vi) process compliance and verification. Experimentation revealed that not only product quality but also process quality varies significantly among student teams and mining process aspects can help the instructor in giving directed and specific feedback. 

\subsubsection{Mining developers' behavior}

An investigation on how developers spend their time based on a fine-grained IDE interaction events dataset is presented in \cite{Minelli2015ITime}. Its authors propose an inference model of development activities to precisely measure the time spent in editing, navigating and searching for artifacts, interacting with the UI, and performing corollary activities, such as inspection and debugging. 

In \cite{Salza2018DoApps}, the authors present an empirical study where app stores were mined to find out if developers update third-party libraries in mobile apps and also to identify update patterns. Evidence found unveiled that mobile developers rarely update their apps regarding used libraries and when they do, they mainly update GUI-related ones.

The measurement of developers' elapsed time in program comprehension activities beyond their IDE interactions is described in \cite{Xia2018MeasuringProfessionals} in a field study with professionals. Findings showed that, on average, developers spend 58\% of their time on program comprehension activities, and they frequently use web browsers and document editors to perform program comprehension activities. Regarding the impact of programming languages, developers' experience, and project phase on the time spent on program comprehension, evidences shown that senior developers spend significantly fewer percentages of time on program comprehension than junior developers.

The assessment of development behaviors and testing practices in real-world projects is reported in \cite{Beller2019DeveloperBehavior}. The authors performed a study involving thousands of developers who were monitored closely on their development activities during the usage of four different IDEs. Results demonstrated that half of the developers' population does not test programs and they rarely run their tests in the IDE. Regarding the behaviors and beliefs towards Test-Driven Development (TDD), findings show this activity as a nonfrequent practice, and software developers only spend a quarter of their time engineering tests, whereas they think they test half of their time.

\subsubsection{Mining end-users' behavior}

Guidelines for the analysis of data collected during software in operation (i.e. when a software product is used by its end-users) are presented in \cite{Pachidi2014}. The authors adopted techniques for extracting knowledge on software operation data, such as users' profiling, clickstream analysis, and classification analysis.

\subsection{Wrap-up}

The aforementioned approaches use a series of n-gram models, topic modeling, and process mining methods mainly to assist programmers in their most basic daily duties, and to discover how end-users operate software products.

Our work uses similar methods however, it focuses on finding developers' fingerprints with the aim of understanding and profiling programmers' behaviors. This approach may provide professors a way to assess students' performance within class tasks. Regarding software and project managers, at an enterprise level, they may use it to improve their task assignment strategies depending on project characteristics, devising adequate replacements in turnover situations, and balancing the constitution of software teams. As for process quality monitoring and enhancement, it can help in finding the good and bad processes followed by a development team or organization.


\section{Study Setup}
\label{sec:study-setup5}

In a controlled experiment where the main objective is analysing programmers' behavior, there is an obvious main source of variability that should be blocked: the nature of the programming task itself. In other words, the optimal setting is to have several programmers\footnote{as many as possible to achieve statistical significance} performing the same task. Other sources of variability are the programming language used, the IDE used, the working conditions and available schedule. 

Being able ``recruit" participants in industry for such an experiment, while blocking all the aforementioned factors is not feasible in a professional context. However, we were able to do that during an academic event dubbed Pythacon\footnote{A twisted contraction of Python + Hackaton : \href[pdfnewwindow=true]{https://sites.google.com/iscte-iul.pt/pythacon}{https://sites.google.com/iscte-iul.pt/pythacon}}. In this event, the same well-defined tasks on software development were performed individually by many participants. Pythacon's first phase consisted of taking a Python In class-MOOC \cite{Gomes2020LearningToCode}. The second phase consisted in a programming contest with six problems with increasing difficulty. The Mooshak\footnote{Available from its home page at \href[pdfnewwindow=true]{http://www.ncc.up.pt/mooshak}{http://www.ncc.up.pt/mooshak}} automatic judge was used to assess participants' performance in their quest for producing solutions in Python for the aforementioned problems \cite{Leal2003Mooshak:System}.

The subjects of this experiment were undergraduate students from three 1st cycle Bologna degrees\footnote{LEI (Computer Engineering), ETI (Telecommunications and Computer Engineering), IGE (Computer Science and Business Management) and LCD (Data Science)} at \href[pdfnewwindow=true]{https://www.iscte-iul.pt}{Iscte}, a public university in Lisbon, Portugal. LCD students did not attend the first phase because their syllabus already included two courses on Python. As such, they acted as the control group regarding the ``treatment'' of taking an in-class MOOC.

All subjects acted as Python developers while trying to build solutions to the proposed problems, upon the PyCharm IDE\footnote{\href[pdfnewwindow=true]{https://www.jetbrains.com/pycharm/}{https://www.jetbrains.com/pycharm/}}, in the same premises\footnote{A large open-space where each participant had an individual table, a portable computer and good natural light} and an equal sprint duration (4 hours). As such, the aforementioned confounding factors were blocked. 
We developed a PyCharm plugin that captures all relevant IDE events, such as, navigational, editing and debugging actions. Each Pythacon participant installed it in its IDE right after reading and signing an informed consent. When starting the IDE for the first time after plugin installation, they were requested to provide their student id number, that was added to the events log. As for the Mooshak automatic judge, that somehow mimics a continuous integration pipeline with an acceptance test battery, it has embedded login and logging mechanisms that allow identifying each participant and its events (problem submissions and corresponding outcomes).

\subsection{Development sessions extraction and storage}

Interaction events collected with our PyCharm plugin, were stored in a \texttt{JSON} file on each subject's computer. A sample event instance is presented in Listing \ref{lst:eclipse_event}. The field tags are self-explanatory.

By the end of Pythacon's programming contest, all event files were uploaded to a central server. Data were then stored into a MySQL database table where the username and event timestamp were composed as an unique key for purging duplicated data. The BPMN model in Figure \ref{chapter3-0-1.pdf} presents the complete schema for the data collection workflow.


\newpage

\lstset{
    xleftmargin=\parindent,
    string=[s]{"}{"}, stringstyle=\color{black},
    comment=[l]{:},  commentstyle=\color{iscte-iul-palette}
}
\begin{lstlisting}[basicstyle=\tiny, caption=Sample PyCharm Event Instance, label=lst:eclipse_event]
{ 
"session"  : "c51973e3-562a-4b65-b6df-49f4c37792e1",
"timestamp_begin" : "2020-09-18T09:00:06.054Z",
"username"  : "87788",
"graduation" : "IGE",
"projectname"  : "PythaconResolution",
"filename"  : "P4.py",
"extension"  : "py",
"categoryName": "NavBarToolbar",
"commandName": "Run",
"platform": "JetBrains s.r.o. / PyCharmCore",
"platform_branch": "PyCharm",
"platform_version": "2020.2.1",
"java": "11.0.8+10-b944.31",
"os": "Mac OS X 10.15.6",
"os_arch": "x86_64",
"country": "Portugal",
"city": "Lisbon",
....
"hash": "0000a3a2cf78485419f15d7913789b16" //To detect event tampering
}
\end{lstlisting}

\plot{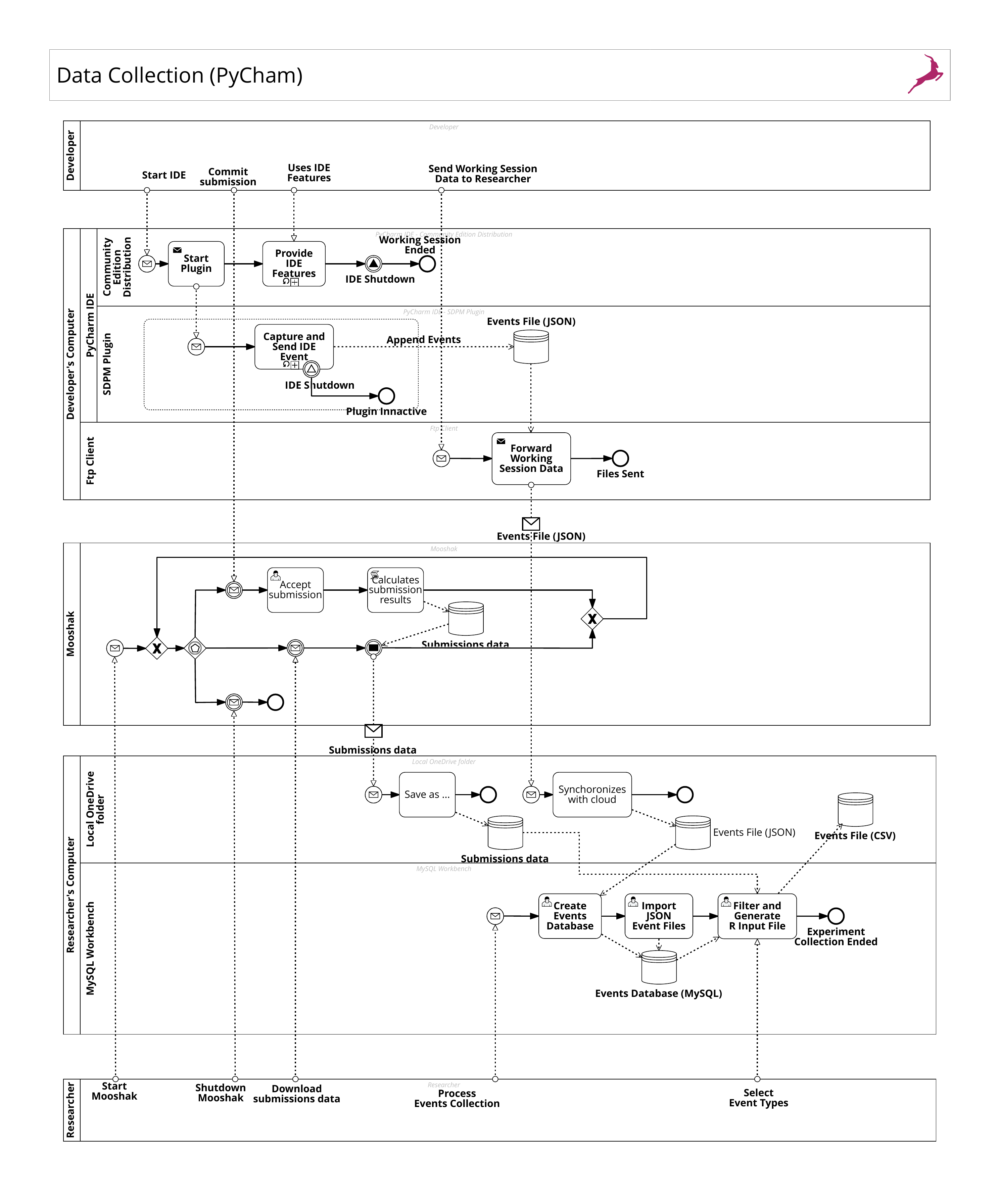}[Data Collection Workflow][13][ht!][trim=1.9cm 1.5cm 1.9cm 3.6cm]

\subsection{Data analysis}
\label{data_analysis}


The complete workflow followed in data pre-processing, aggregation and analysis is presented in Figure \ref{chapter3-0-2.pdf}.

\plot{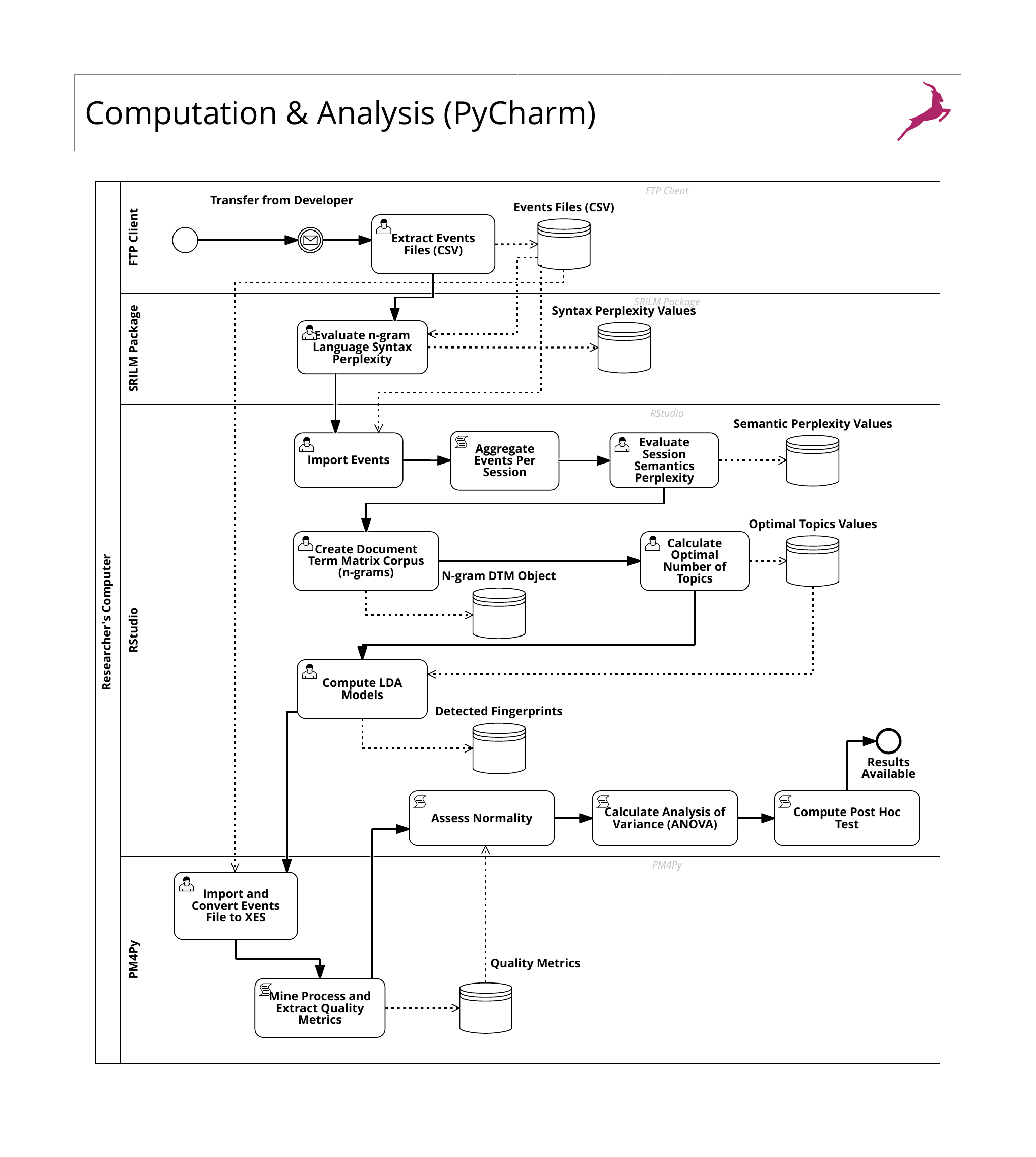}[Study Computation and Analysis Process][13][ht!][trim=1.9cm 1.8cm 1.9cm 3.6cm]

\subsubsection{N-gram language models evaluation}

Documents containing natural language, software code or development sessions, are often repetitive and highly predictable. A good language model should capture the regularities in the corpus. If carefully produced from a representative corpus, it will predict, with high confidence, the contents of a new document drawn from the same population. In other words, the model will not find a new document particularly surprising. In natural language processing (NLP), this idea is captured by a measure called perplexity, or its log-transformed version, cross-entropy \cite{Hindle2012OnSoftware}.

Given a document containing the textual representation of a development session within the IDE,
$s = a_1, \dots, a_n$, where terms represent development commands or activities,
and a language model $M$, we assume that the probability of the document estimated by the model is $p\textsubscript{M}(s)$. We can write down the cross-entropy measure as:
\begin{equation}
    H_M(s) = -\frac{1}{n}\log p_M(a_1, \dots, a_n)
\end{equation}

and by the formulation presented in earlier:
\begin{equation}
    H_M(s) = -\frac{1}{n}\sum_1^n \log p_M(a_i|a_1, \dots, a_n)
\end{equation}

This measures how ``surprised'' a model is by looking at an unseen document. A model with low entropy for target documents is expected to be a good model. Higher probabilities are given (closer to 1, and thus lower absolute log values) to more frequent words, and lower probabilities to rare ones. 
If a hypothetical optimal model is deployed to predict developers' actions, it is possible to guess what the next action would be and at the same time characterize developers' behaviors.

To shed light on how regular development sessions are, we performed a series of experiments with both natural language and development sessions corpora, first comparing the ``naturalness'' (using cross-entropy) of IDE actions on development sessions with English texts, and then comparing various session corpora to each other to further gain insight into the similarities and differences between sessions corpora.

Our natural language studies were based on a R package with widely used corpora from Jane Austen's novels\footnote{\href[pdfnewwindow=true]{https://cran.r-project.org/web/packages/janeaustenr/index.html}{https://cran.r-project.org/web/packages/janeaustenr/index.html}}.
To compute the models perplexity and obtain the correspondent cross-entropy, we used the SRILM package\footnote{SRILM Toolkit - \href[pdfnewwindow=true]{http://www.speech.sri.com/projects/srilm}{http://www.speech.sri.com/projects/srilm}}.
All the models were evaluated using a 5-fold cross validation strategy, meaning the corpus was randomly divided into 5 parts, where 80\% was used as the training set and 20\% as the test set, and this process was repeated 5 times.

\subsubsection{Topic models evaluation}
\label{topics-evaluation}

To determine the optimal number of topics to model developers' sessions, we used the R package \textit{ldatuning}\footnote{\href[pdfnewwindow=true]{https://cran.r-project.org/web/packages/ldatuning/ldatuning.pdf}{https://cran.r-project.org/web/packages/ldatuning/ldatuning.pdf}}, that applies an empirical approach rather than intuition.
Metrics such as CaoJuan2009 \cite{Cao2009ASelection} and Arun2010 \cite{Arun2010OnObservations} are to be minimized (tend to 0), whilst metrics like Deveaud2014 \cite{Deveaud2014AccurateRetrieval} and Griffiths2004 \cite{Griffiths2004FindingTopics} are expected to be maximized (tend to 1). The lower the distances to the objective values, the better the model and, consequently, the optimal number of topics are found in that particular point.

\subsubsection{Process models evaluation}
\label{sec:process-metrics-evaluation}

Process Mining is now a mature discipline with validated techniques producing accurate outcomes on several business domains \cite{Poncin2011a}. Discovery is the ability to construct a process model, by capturing the behavior of a process based on an event log \cite{VanderAalst2016}.



Following model discovery, conformance checking stands for the confrontation of a process model with the ``reality'' represented by the logged events during the actual execution of the corresponding deployed process. Conformance checking can be used to detect deviations from prescribed processes, determine differences and/or similarities between process variants or verify the accuracy of documented processes \cite{VanderAalst2016}. It can also be used to calculate the efficiency or to measure the quality of a process model. Quality is normally assessed considering four metrics:
\begin{itemize}
    \item \textbf{Fitness.} Represents how much behavior in a log is correctly captured (or can be reproduced) by a discovered model \cite{Berti2020AEnhancement}.
    \item  \textbf{Precision.} Refers to how much more behavior is captured in the model than what was observed in the log. It deals with avoiding overly under fitted models \cite{Munoz-Gama2010AConformance}.
    \item  \textbf{Generalization.} Focuses on avoiding overly precise models based on the assumption that logs are by their nature incomplete, meaning that, to a certain extent, a model should be able to reproduce not yet seen behaviour in the log \cite{Buijs2014QualitySimplicity}.
    \item \textbf{Simplicity.} Alludes to the rule that the simplest model that can describe the behavior found in a log is indeed the best model. Model complexity, the opposite of simplicity, is dependent on the number of nodes and arcs in the underlying graph \cite{RojasBlum2015MetricsDiscovery}.
\end{itemize}

To calculate the previous metrics we used the Process Mining library for Python (PM4Py)\footnote{\href[pdfnewwindow=true]{https://pm4py.fit.fraunhofer.de/documentation\#discovery}{https://pm4py.fit.fraunhofer.de/documentation\#discovery}}.



\subsection{Research questions}

The research questions for this work are the following:

\begin{itemize}

\item \textbf{RQ1:} Do n-gram language models capture local regularities in software development sessions?\\
\textbf{Methods used.} Computation of n-gram language models perplexity/cross-entropy using SRILM and LDA with n-gram windows.

\item \textbf{RQ2:} Can we coherently characterize development sessions in terms of fingerprints?\\
\textbf{Methods used.} Topic Modeling using the LDA algorithm with n-gram window tuning.

\item \textbf{RQ3:} Are there any significant variation in sessions simplicity and interactions magnitude between distinct participants?\\
\textbf{Methods used.} Process models discovery using the Directly Follows Graph mining algorithm and hypothesis testing.


\end{itemize}

\section{Study Results}
\label{sec:study-results5}
In this section, we present the results of our experiment, regarding its research questions.

\begin{table}[H]
\scriptsize
\caption{Participants Statistics}
\label{table:participants-statistics}
\centering
\begin{tabular}{p{1.5cm}ccccc}
	\hline\noalign{\smallskip}
			\textbf{} & \textbf{LEI} & \textbf{ETI} & \textbf{IGE} & \textbf{LCD} & \textbf{Total}\\
	\noalign{\smallskip}\hline\noalign{\smallskip}
	
    Participants & 12 & 9 & 7 & 9 & 37\\[0.1cm]
    
    Attended MOOC & Yes & Yes & Yes & No & 28 \\[0.1cm]
    
    
   \noalign{\smallskip}\hline\\
\end{tabular}
\end{table}


\subsection{\textbf{RQ1. Do n-gram language models capture local regularities in software development sessions ?}}

\subsubsection{Local syntactic structure}
\label{syntax-evaluation}

To answer this question we estimated n-gram models of plain English corpus and the development session IDE commands and their categories.

From Figure \ref{chapter5-9-1.pdf}, we observe that, although English has a higher level of cross-entropy across all n-gram models, it declines rapidly, saturating around tri- or 4-grams. The same happens with our development sessions models, which have generally lower cross-complexity for unigram models, and also saturate around tri-grams models.
This indicates, as expected, that development sessions repetitive context can also be captured by language models. 

\plot{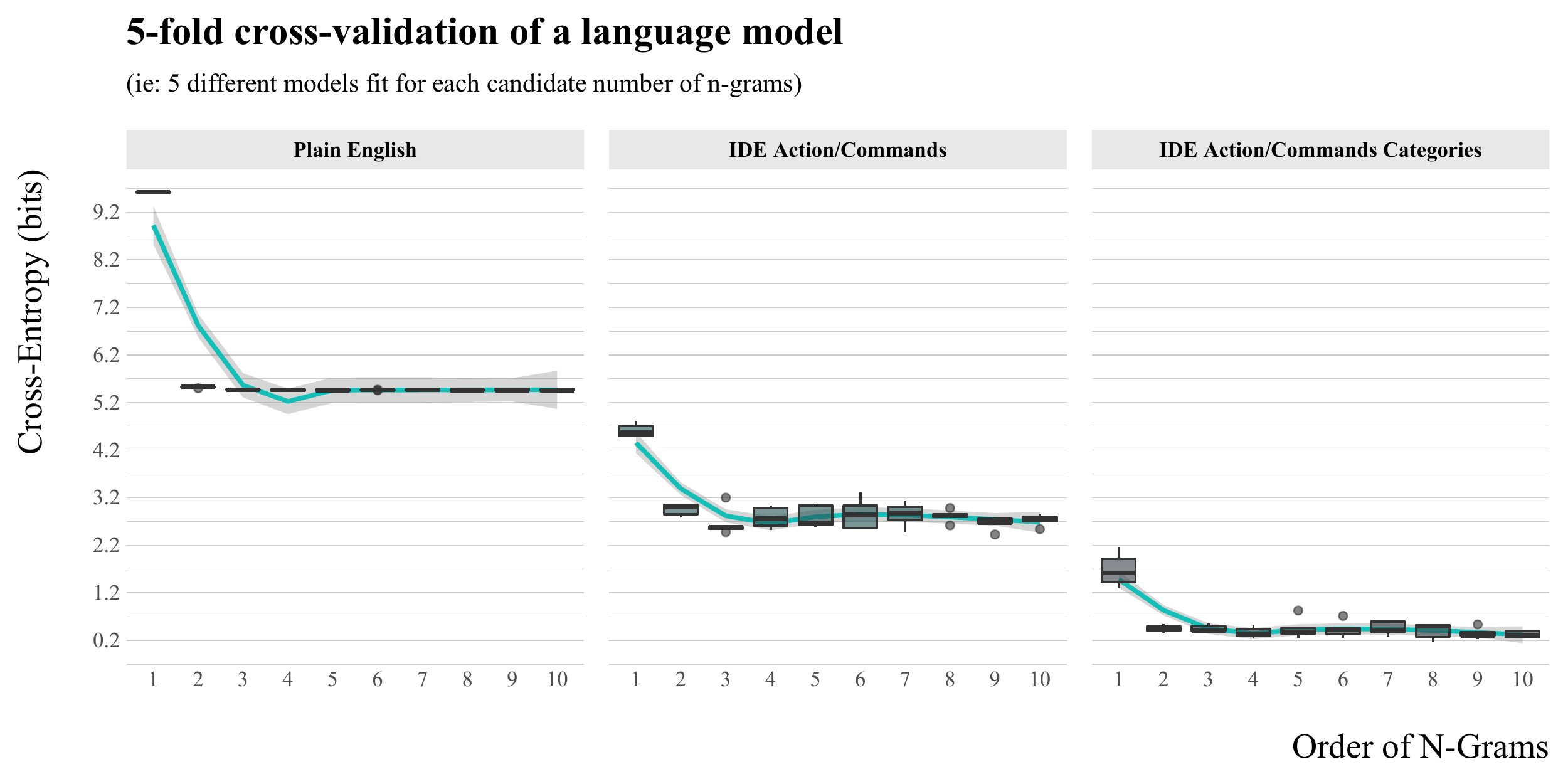}[Plain English vs. Python Development Sessions Cross-Entropies using n-gram models][13.8][ht!][trim=0cm 0cm 0cm 0cm]


We find that a typical development session is far more regular than English, with entropies starting from 4.2 bits and declining to 2.7 bits by IDE command and starting at 2.2 bits and saturating around 0.7 bits for command categories. 

Our findings may have implications in the way we manage developers' activities. They provide more and detailed evidence to confirm what was already mentioned by \cite{Damevski2018PredictingModels}, the possibility to design and build even more optimized recommendation systems to help and guide developers on the activities they are executing or should be doing next. Moreover, they shed light on the optimal number of n-grams to use, thus avoiding the waste of computing resources and at the same time provide further evidence for the usefulness of using text mining techniques to detect and monitor developers' behaviors.





\subsubsection{Semantics}
\label{semantic-evaluation}

In the context of IDE usage, each development session may have its own semantics. Whilst to capture the local syntactic structure of a language we used n-gram language models, to assess the semantics of the development sessions we used LDA. Figure \ref{chapter5-15-3.pdf} shows the cross-entropy regarding the semantics analysis for the development sessions. It consists in finding the entropy for n-gram models, each having $k$ topics, where $k$ varies from 2 to 10, and where each combination of n-grams and $k$ topics was calculated with a 5-fold cross-validation strategy.


 \plot{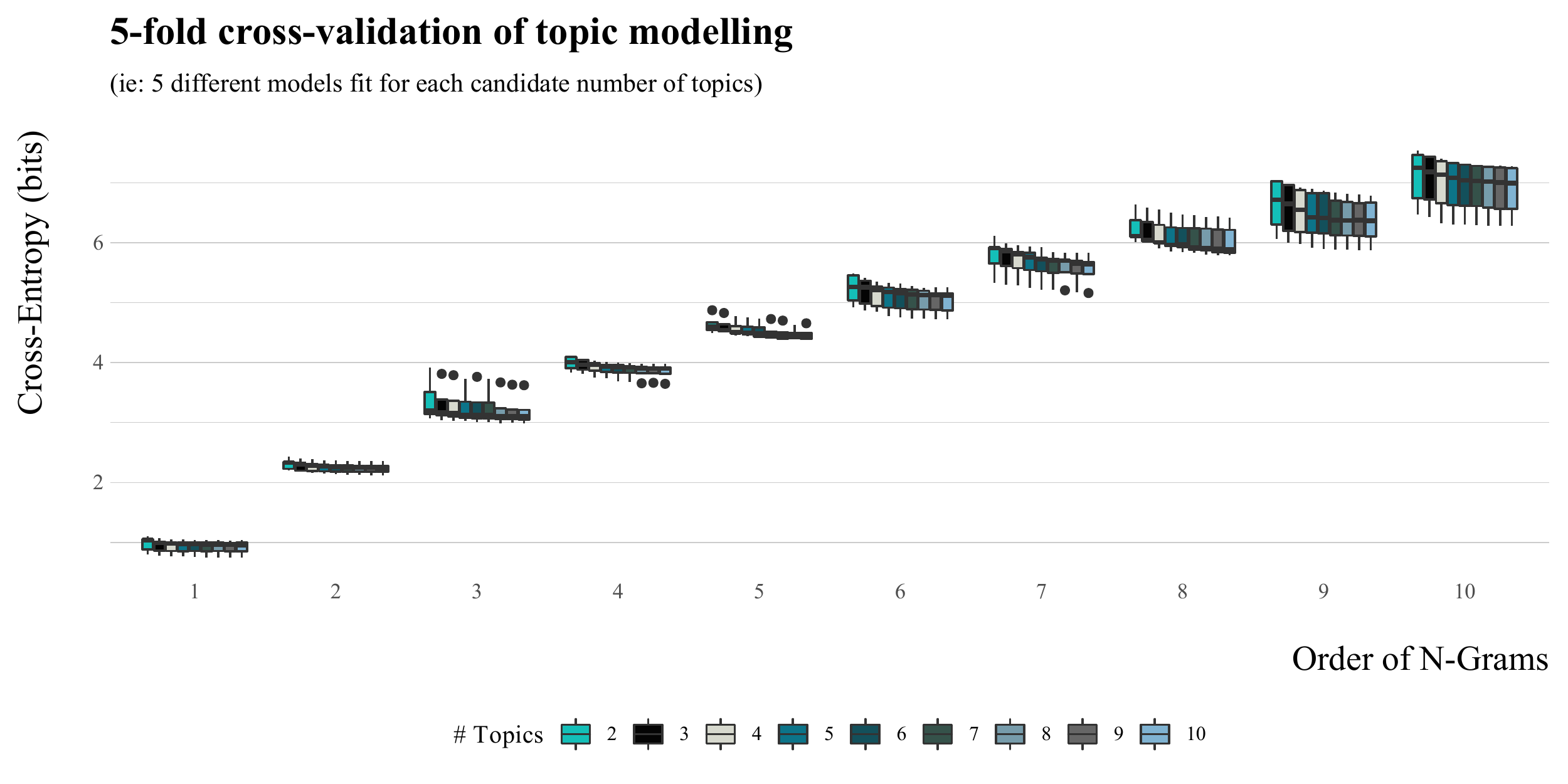}[Development sessions modeling using LDA with $k$ topics and n-grams][13.8][ht!][trim=0cm 0cm 0cm 0cm]
 
As one can easily observe, when using LDA to assess the number of topics, entropy grows with the order of the n-grams. The higher the n-gram window the less the model is able to predict future cases because the perplexity is higher. 
Regarding the number of topics on each n-gram model, we can confirm the expected behavior, when the number of topics increase, independently of the n-gram model, the entropy tends to decline.

Concerning the interpretation of the n-gram results perspective, they show that, given the randomness of IDE actions performed by developers, to increase the n-gram value in characterizing a session, decreases the ability for LDA to find similar ones. As for the number of topics, the bigger the number of topics the better the model can detect similar sessions. 

Based on Figure \ref{chapter5-15-3.pdf}, we argue that, when using LDA to detect similarities within development sessions, we should evaluate carefully the use of more than tri- or four-gram models. In one hand we know that the higher the n-gram model, the higher the computational resources needed. On the other hand we have evidences that five-gram models have an entropy of around 5 bits, which is by itself a high value.


\subsection{\textbf{RQ2. Can we coherently characterize development sessions in terms of fingerprints ?}}

Figure \ref{chapter7-1.pdf} in section \ref{language-models} provided a small sample of the activities aggregating the commands issued by developers. Those were defined according to the method used by \cite{Minelli2015ITime}, and by adding extra activities reflecting the results of the submission events. Regarding IDE interactions, the commands were recoded into activities like: \textit{Editing, Navigating, Debugging, Refactoring, Executing and Spurious}. As for the submission actions, we used their native identifiers : \textit{Accepted\_Answer, Wrong\_Answer, Compile\_Time\_Error, Invalid\_Submission, Runtime\_Error and Time\_Limit\_Exceeded}. From these, we computed the optimal number of patterns(topics) by assessing the probabilistic coherence of multiple topics using the metrics described in \ref{topics-evaluation} and uni-gram, bi-grams and tri-grams models only.

\plot{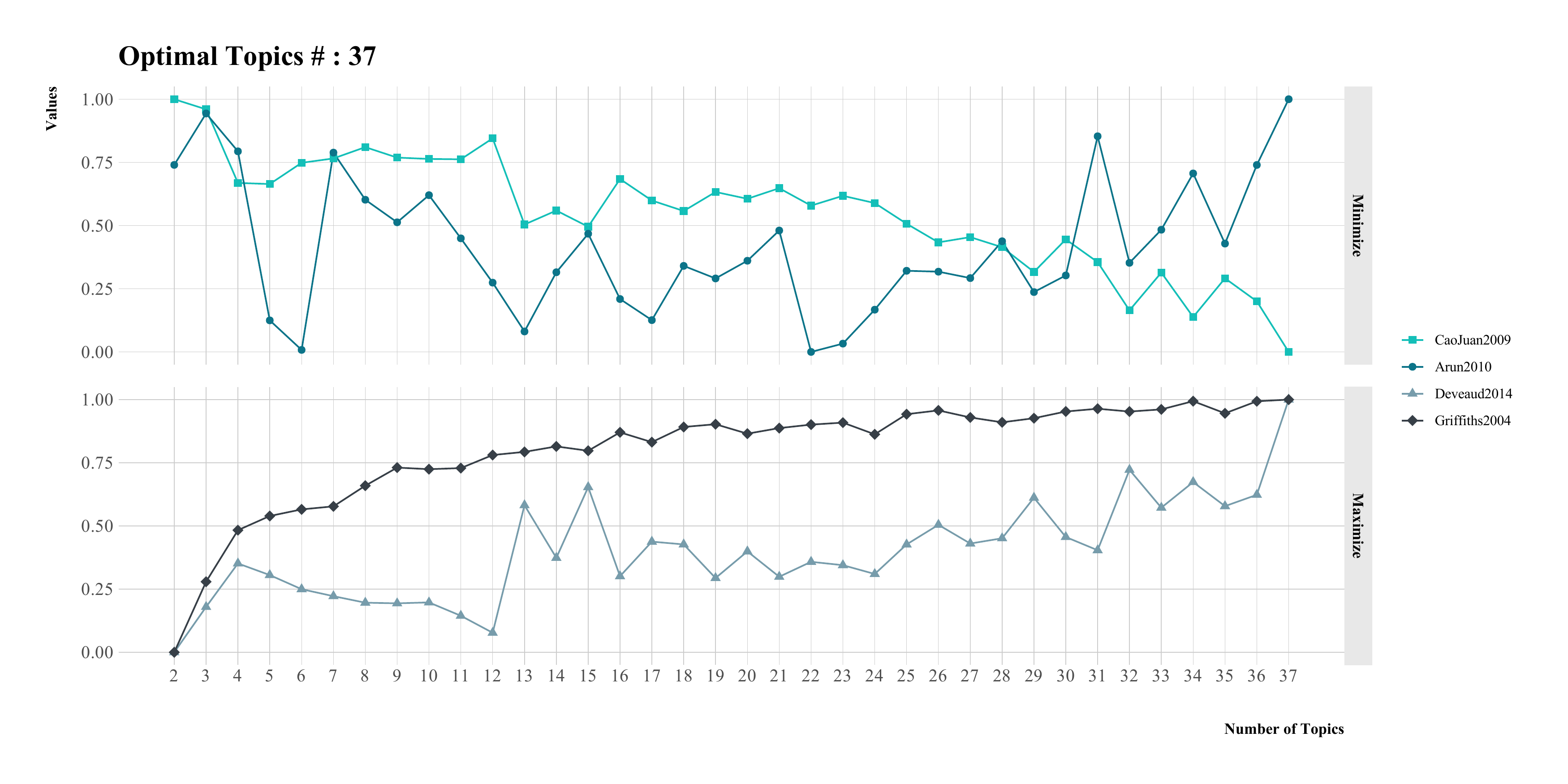}[Assessing the optimal number of distinct patterns to search][14][ht!][trim=0cm 0cm 0cm 0cm]

\subsubsection{\textbf{Optimal number of patterns}}


To decide the optimal number of patterns, we took into account the highest value for the number of topics where any of the metric is close to the objective value, either when minimizing or maximizing its value and therefore, we picked 37. This number represents the size of the population, which in reality confirms that there is a great deal of variance between sessions.
When applying the LDA algorithm with k=37, due to the average of the probabilities of an activity to belong to a session and the average of probabilities of a participant to belong to a specific session pattern, LDA has placed the participants in only 19 different patterns. Figure \ref{chapter7-0.pdf} shows the optimal number of topics evaluation.

\subsubsection{\textbf{High performers}}

Figure \ref{chapter7-6.pdf} shows the topics, or in our case, the referred fingerprints, identified to characterize what we call the high performers\footnote{Eight participants were in this condition.}, the same is to say, the ones with good process smells. In this group we have those who ranked above quartile 3, meaning that they have answered correctly to at least four exercises. From the six fingerprints found for those participants, we can observe the following:

\begin{itemize}

\item \textbf{Fingerprint 19. Cautious coder. Aggressive executor.} Contains a profile of development centered in frequent editing actions, mixed with permanent execution of the code to validate the result before submitting to Mooshak. This pattern was the fingerprint of participant D.

\item \textbf{Fingerprint 20. Cautious coder.} Reveals a similar pattern, however, with less prevalence of program execution and therefore less testing actions. Exercise submission actions are very rare. With this fingerprint we find 2 participants, G and H.

\item \textbf{Fingerprint 24. Cautious coder. Test skipper.} With no surprise, in this fingerprint, we find editing also as the most frequent action. However, the next common action is not program execution, but submissions for validation. This pattern was the fingerprint of participant C.

\item \textbf{Fingerprint 26. Insecure. Testers.} Participants characterized in this group followed explicitly a permanent program execution, followed by editing activities. They have submitted their answers infrequently, meaning that they have probably tested well their work before any submission. This pattern was the fingerprint of participant F.

\item \textbf{Fingerprint 30. Insecure. Debuggers.} This pattern reveals participants more focused on debugging activities, followed by a mix of editing and navigational actions. In a certain sense, it looks as if they have replaced their program execution tests with fine grain debugging practices. This pattern was the fingerprint of participant E.

\item \textbf{Fingerprint 35. Balanced coders. Confident.} Provides evidence for a pattern of high frequency in editing, followed by a balanced persistence of program execution practices and navigational activities.  There were however no frequent activities related with the submission of code to answer the exercises. It suggests these participants only submitted their answers after careful review of their code and without the need for deeper debugging tasks. With this fingerprint we find the top 2 participants, A and B.

\end{itemize}

\plot{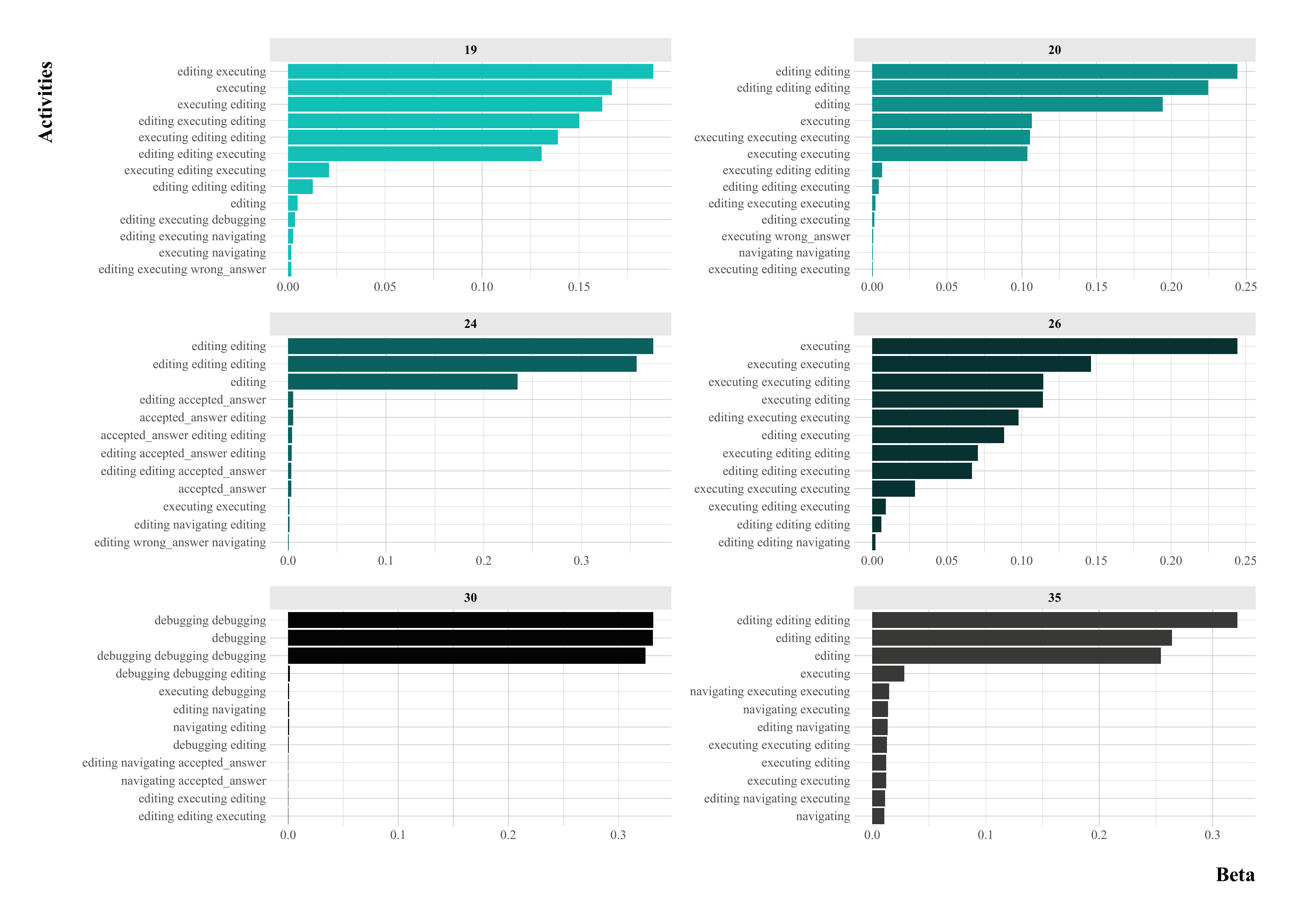}[Development fingerprints for the top(8) performers][13][ht!][trim=0cm 0cm 1cm 1cm]


\subsubsection{\textbf{Low performers}}

Figure \ref{chapter7-7.pdf} represents the characteristics of those who had more difficulties in executing the tasks and which we may consider as having bad process smells. It plots the unique fingerprints of the last eight participants, the ones with zero or just one correct answer.

\begin{itemize}

\item \textbf{Fingerprint 7. General coding limitations.} Reveals a practice focused almost exclusively on editing actions, and a very small prevalence (near zero) of navigational, program executions or exercise submission activities. This pattern was the fingerprint of participant V.

\item \textbf{Fingerprint 18. General coding limitations.} Shows a similar profile as fingerprint 7 regarding editing practices. However, program executions and answer submissions appear more often in the complete work sessions, yet with a low frequency (near zero) when compared with editing. This pattern was the fingerprint of participant U.

\item \textbf{Fingerprint 25. Limited python/algorithmic skills.} It characterizes a practice where editing is also the prevalent action and combines this frequently with debugging and program execution activities. Answer submission is however infrequent, as none shows in the most common actions. This pattern was the fingerprint of 5 participants, S, T, W, X and Y. None of them has attended the MOOC training sessions.

\item \textbf{Fingerprint 33. Limited python/algorithmic skills.} This practice is characterized as usual by frequent editing actions, and then followed by a decreasing and balanced editing, navigational and refactoring decisions. However, submission actions are absent. This pattern was the fingerprint of participant Z.

\end{itemize}

\plot{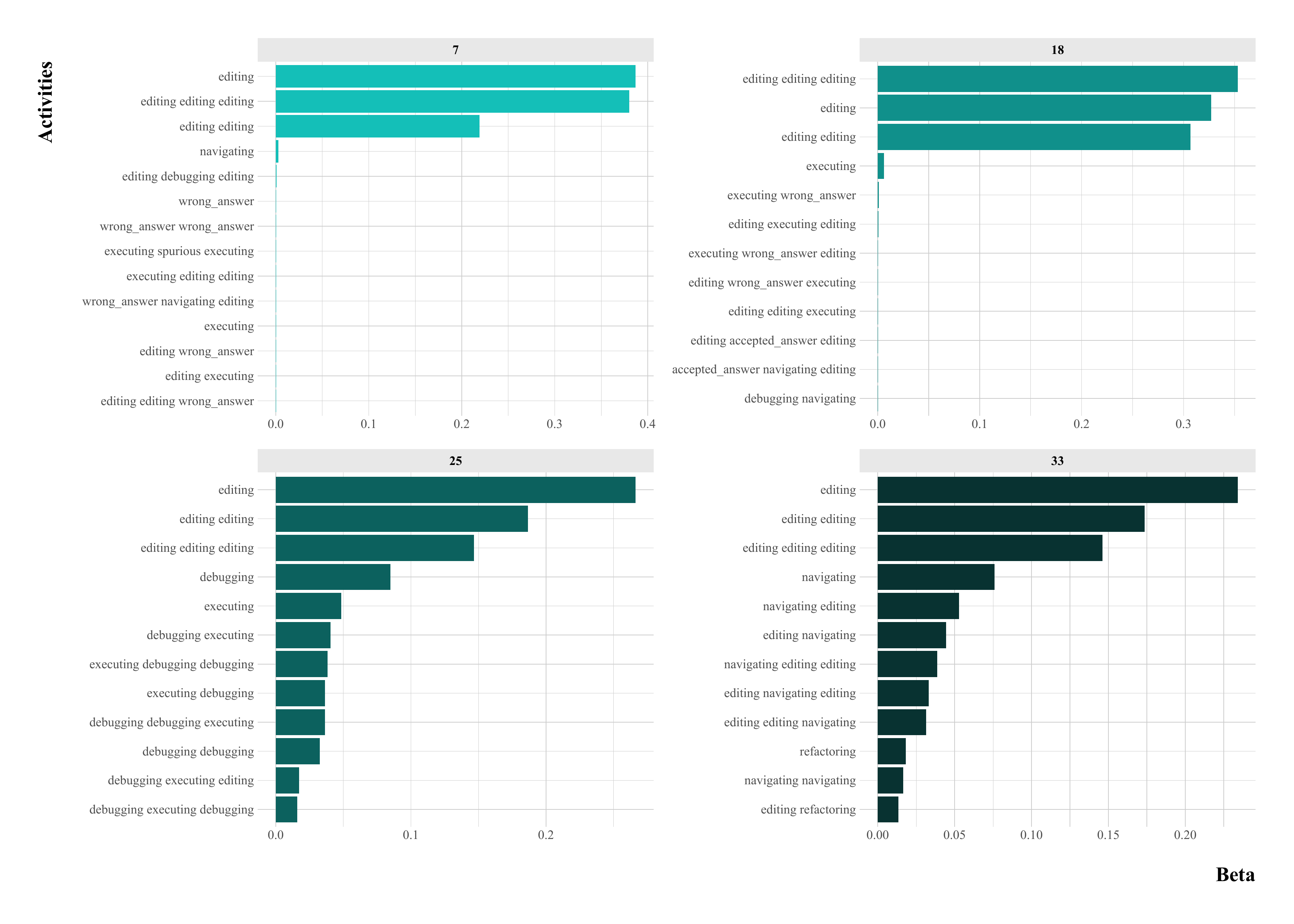}[Development fingerprints for the bottom(8) performers][13][ht!][trim=0cm 0cm 1cm 1cm]


\plot{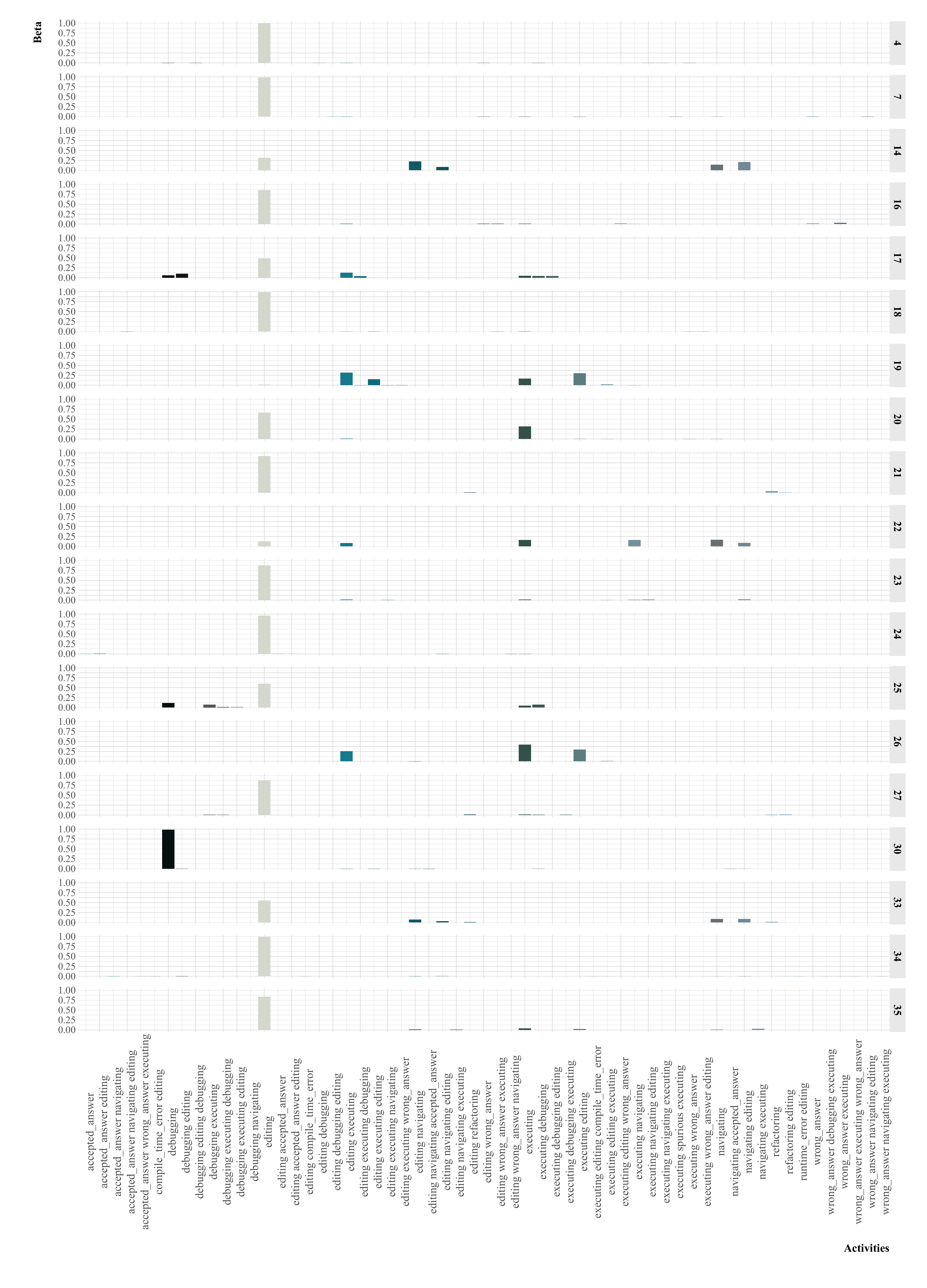}[Development fingerprints characterizing all participants][14][ht!][trim=1cm 1.5cm 1cm 0cm]

It is striking to observe that the fingerprints characterizing the high performers are significantly different, either in the top activities and also in probabilities, from the ones describing the low performers.
Figure \ref{chapter7-20.pdf} presents the distinct fingerprints detected to characterize all participants. Based on these findings, one may argue that the variation in the participants scores was only due to the quality of the code they have produced, moreover, that the variation in the fingerprints was due to their own programming skills. Additionally one may suggest as an explanation for the top performers, the knowledge acquired in the course they belong to, due to the MOOC training attendance or the dimension of their coding interactions during the contest and not based on their coding behaviors.
That is a possibility we cannot reject immediately. However, we may assess this hypothesis if we mine, from a different perspective, the overall process for each participant, course as a group of participants with same backgrounds, according to their MOOC training participation and finally according to their performance. If the reason for the higher performance is related with the magnitude of their interactions or the quality of their code, we expect to see no significant variance in the process simplicity amongst different participants. On the contrary, if variation in the process exists between groups, that may be an indication that the quality of the outcome is indeed related to the development workflow.
Following the above rationale, we mined the correspondent processes using a mining algorithm appropriate for processes with thousands of events and where a fuzzy or spaghetti-like process behavior is expected to exist. We used a Directly-Follows Graph algorithm from the Process Mining library for Python mentioned earlier, and assessed the  models produced using the quality metrics described in section \ref{sec:process-metrics-evaluation}.
Table \ref{table:process-models-evaluation} summarizes the fingerprint results for the referred participants, along with the metrics used to evaluate the quality for the process models discovered for each of them. The hypothesis we later tested was as follows:
Are there significant differences in the processes complexity or development interactions between the different graduation courses or between the top, bottom and the rest of the participants ?.

\begin{landscape}

\begin{table}[ht!]
\tiny
\caption{Process Models Evaluation}
\label{table:process-models-evaluation}
\centering
\begin{tabular}{p{1.8cm}ccccccccc}
	\hline\noalign{\smallskip}
			\textbf{} & \textbf{Course}  & \textbf{Fingerprint} & \textbf{Interactions} & \textbf{Fitness} & \textbf{Precision} & \textbf{Generalization} & \textbf{Simplicity} & \textbf{Average} & \textbf{Duration}\\
	\noalign{\smallskip}\hline\noalign{\smallskip}
	
	\multicolumn{10}{l}{\cellcolor{gray!10}\textbf{By Course}}\\[0.1cm]
    LEI & -- & -- & 33011 & 0.017 & 1 & 0.142 & 0.432 & 0.398 & 00:07:09\\[0.1cm]
    ETI & -- & -- & 26557 & 0.818 & 1 & 0.153 & 0.439 & 0.602 & 00:05:01\\[0.1cm]
    IGE & -- & -- & 16635 & 0.199 & 1 & 0.143 & 0.438 & 0.445 & 00:02:19\\[0.1cm]
    LCD & -- & -- & 30057 & 0.198 & 1 & 0.126 & 0.426 & 0.438 & 00:06:47\\[0.1cm]
 
	\multicolumn{10}{l}{\cellcolor{gray!10}\textbf{MOOC Training}}\\[0.1cm]
	MOOC & -- & -- & 76203 & 0.039 & 1 & 0.124 & 0.411 & 0.394 & 00:27:27\\[0.1cm]
    NO\_MOOC & -- & -- & 30057 & 0.198 & 1 & 0.126 & 0.426 & 0.438 & 00:06:18\\[0.1cm]

	\multicolumn{10}{l}{\cellcolor{gray!10}\textbf{Performance Type}}\\[0.1cm]
    High (Top 8) & -- & -- & 23351 & 0.009 & 1 & 0.140 & 0.437 & 0.396 & 00:04:07\\[0.1cm]
    Low (Bottom 8) & -- & -- & 25869 & 0.957 & 1 & 0.153 & 0.447 & 0.639 & 00:04:10\\[0.1cm]
    Middle & -- & -- & 57040 & 0.037 & 1 & 0.121 & 0.410 & 0.392 & 00:19:58\\[0.1cm]
    
    \multicolumn{10}{l}{\cellcolor{gray!10}\textbf{High Performers}}\\[0.1cm]
    A & ETI & 35 & 4447 & 0.089 & 1 & 0.158 & 0.489 & 0.434 & 00:00:32\\[0.1cm]
    B & IGE & 35 & 3554 & 0.123 & 1 & 0.181 & 0.538 & 0.460 & 00:00:21\\[0.1cm]
    C & LEI & 24 & 1515 & 0.908 & 1 & 0.192 & 0.507 & 0.652 & 00:00:05\\[0.1cm]
    D & LEI & 19 & 2194 & 0.118 & 1 & 0.202 & 0.524 & 0.461 & 00:00:09\\[0.1cm]
    E & IGE & 30 & 4809 & 0.353 & 1 & 0.162 & 0.472 & 0.497 & 00:00:35\\[0.1cm]
    F & IGE & 26 & 2495 & 0.228 & 1 & 0.173 & 0.543 & 0.486 & 00:00:12\\[0.1cm]
    G & LEI & 20 & 2481 & 0.978 & 1 & 0.181 & 0.529 & 0.672 & 00:00:10\\[0.1cm]
    H & LEI & 20 & 1856 & 0.170 & 1 & 0.174 & 0.506 & 0.462 & 00:00:07\\[0.1cm]
  

    \multicolumn{10}{l}{\cellcolor{gray!10}\textbf{Low Performers}}\\[0.1cm]
    S\textsuperscript{*} & LCD & 25 & 3520 & 0.042 & 1 & 0.179 & 0.514 & 0.434 & 00:00:22\\[0.1cm]
    T\textsuperscript{*} & LCD & 25 & 4615 & 0.204 & 1 & 0.186 & 0.524 & 0.478 & 00:00:35\\[0.1cm]
    U & LEI & 18 & 2200 & 0.021 & 1 & 0.173 & 0.521 & 0.429 & 00:00:10\\[0.1cm]
    V & IGE & 7 & 1064 & 0.987 & 1 & 0.189 & 0.617 & 0.698 & 00:00:03\\[0.1cm]
    W\textsuperscript{*} & LCD & 25 & 2599 & 0.128 & 1 & 0.188 & 0.544 & 0.465 & 00:00:12\\[0.1cm]
    X\textsuperscript{*} & LCD & 25 & 4393 & 0.336 & 1 & 0.239 & 0.527 & 0.526 & 00:00:32\\[0.1cm]
    Y\textsuperscript{*} & LCD & 25 & 1964 & 0.145 & 1 & 0.203 & 0.585 & 0.483 & 00:00:07\\[0.1cm]
    Z & ETI & 33 & 2781 & 0.831 & 1 & 0.251 & 0.565 & 0.662 & 00:00:13\\[0.1cm]


   \noalign{\smallskip}\hline\\
   \multicolumn{10}{l}{\textbf{Interactions} - Represent actions within the IDE, \textbf{Duration} - Means the time to build/compute the process model}\\
   \multicolumn{10}{l}{\textbf{*} - Participant did not attend the MOOC}
\end{tabular}
\end{table}

\end{landscape}




\subsection{\textbf{RQ3. Are there any significant variation in sessions simplicity and interactions magnitude between distinct participants ?}}

Simplicity is one of the dimensions to analyze a process model, and to calculate it, PM4Py takes into account only a Petri net model. The criteria adopted for calculating simplicity is the inverse arc degree as described in \cite{Munoz-Gama2010AConformance}. Since we mined individual processes, they would represent the behavior simplicity of each participant in the programming exercise.

Interactions magnitude refers to the sum of the number of command actions executed in the IDE plus the submission of answers in the Mooshak platform. In other words, interactions are represented by the events generated during the programming exercise by each individual.


The objective of this test is to assess if there is a relation between the performance along with the sessions simplicity or magnitude of interactions on different sets of participants. For this purpose we tried the analysis of variance (ANOVA).

\textbf{ANOVA Test.} Tests if there are significant different statistics between groups of participants, or the same is to say, helps to figure out if one needs to accept or reject the null hypothesis. A one way ANOVA is used to compare two means from two independent (unrelated) groups using the F-distribution. The null hypothesis for the test is that the two means are equal. Therefore, a significant result means that the two means are unequal.
It has the ability to tell if at least two groups were different from each other, however, it won’t tell which groups were different and by which magnitude. If a test returns a significant f-statistic, then one may need to run an ad hoc test (eg: Tukey HSD) to learn exactly which groups had a difference in means.

\textbf{Tukey HSD ("honestly significant difference" or "honest significant difference").} Is a statistical tool used to determine if the relationship between two sets of data is statistically significant – that is, whether there's a strong chance that an observed numerical change in one value is causally related to an observed change in another value. In other words, the Tukey test is a way to test an experimental hypothesis.

\subsubsection{Variables Assumptions}
The use of ANOVA has several assumptions, such as: i) the dependent variable should be measured at the continuous level or absolute scale; ii) the independent variables should define at least two categorical treatments, that corresponds to the groups to which the participants belong; iii) there should be no significant outliers in the groups since they can have a negative effect on ANOVA; iv) the distribution of the dependent variables should be as normally distributed as possible. Having all other conditions satisfied, we assessed the normality.

\subsubsection{Normality Tests}
To test normality, we may use two well-known tests of normality, namely the Kolmogorov-Smirnov and the Shapiro-Wilk tests. We only considered the Shapiro-Wilk test to assess normality since the latter is more appropriate for small sample sizes ($<$ 50 samples). The results are presented in Table \ref{table:normality-test5}, and from them, we cannot reject the null hypothesis, therefore, we accept that both Simplicity and Interactions are normally distributed justifying the use of ANOVA.

\begin{table}[H]
\scriptsize
\caption{Normality Tests}
\label{table:normality-test5}
\centering
\begin{tabular}{p{1.5cm}cc}
	\hline\noalign{\smallskip}
	& \multicolumn{2}{c}{\cellcolor{gray!10}\textbf{Shapiro-Wilk}}\\
	\textbf{Factor} &
	\textbf{\textit{Statistics(W)}} & \textbf{\textit{Sig.*}}\\
	\noalign{\smallskip}\hline\noalign{\smallskip}

    \textbf{Symplicity}  & 0.94802 &  0.08334 \\[0.1cm]
    \textbf{Interactions} & 0.95760 &  0.16940 \\[0.1cm]



   \noalign{\smallskip}\hline
   	\multicolumn{3}{l}{\textit{*Statistically significant if Sig. $<$ 0.05}}
\end{tabular}
\end{table}

\subsubsection{Findings}

From Table \ref{table:aov5}, we can confirm the significant variance between the ones with less quality in their code(Bottom5) and the rest of the participants (Top5 and Others), and this difference is larger between the top five and bottom five performers. Figures \ref{chapter5-Top5.pdf} and \ref{chapter5-Bottom5.pdf} show the process models discovered for both, respectively.
These results provide evidences to state that, the differences in the proficiency between certain participants are not only related with the coding skills each of them may have.

The process complexity followed by each developer may also influence the outcome, or at least, may be used as a valid indicator for assessing quality between developers.
However, when these groups are configured to contain the top and bottom eight participants in terms of score, that significance is no longer visible within the same levels of confidence ($\alpha <$ 0.05). This reinforces the fact that if significant differences exist, they are most likely and only in the individual behaviors.
As for analyzing the variance between different courses, we found no significant differences, either for the development sessions simplicity as well as for the magnitude of interactions. 

As mentioned earlier, we have assembled a method to capture local regularities and overall structure of development processes, the so called fingerprints. Based on the data we obtained, namely the probabilities of activities in the development sessions, and the metrics from the discovered processes, we may classify those fingerprints as good or bad process smells and start to create a catalog of software development process smells. Later on, models can be built to evaluate automatically if a coding session is following a good or a bad practice and suggest guidance actions to developers.

\plot{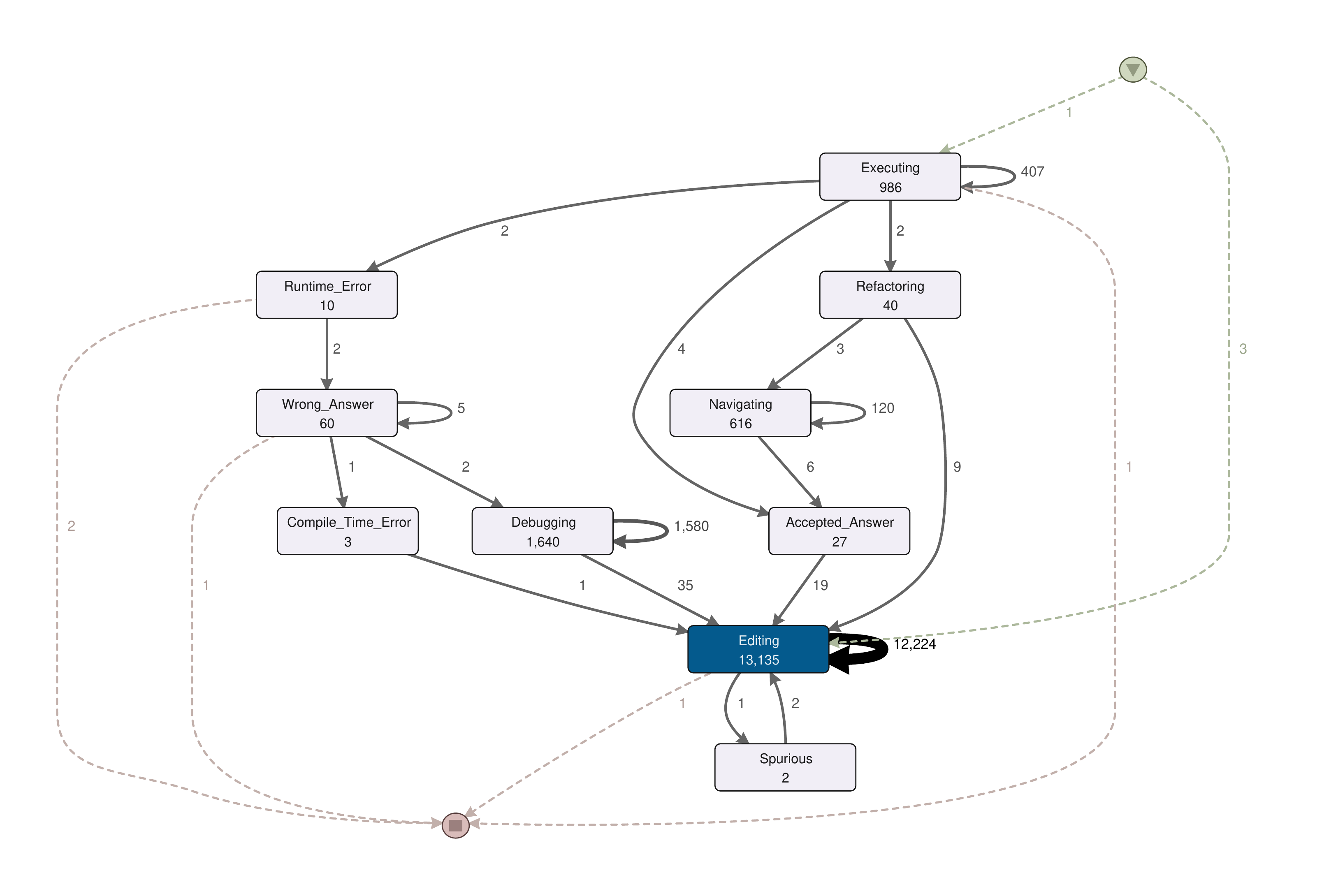}[Process Model characterizing Top5 Participants][14][ht!][trim=1cm 1cm 1cm 0cm]

\plot{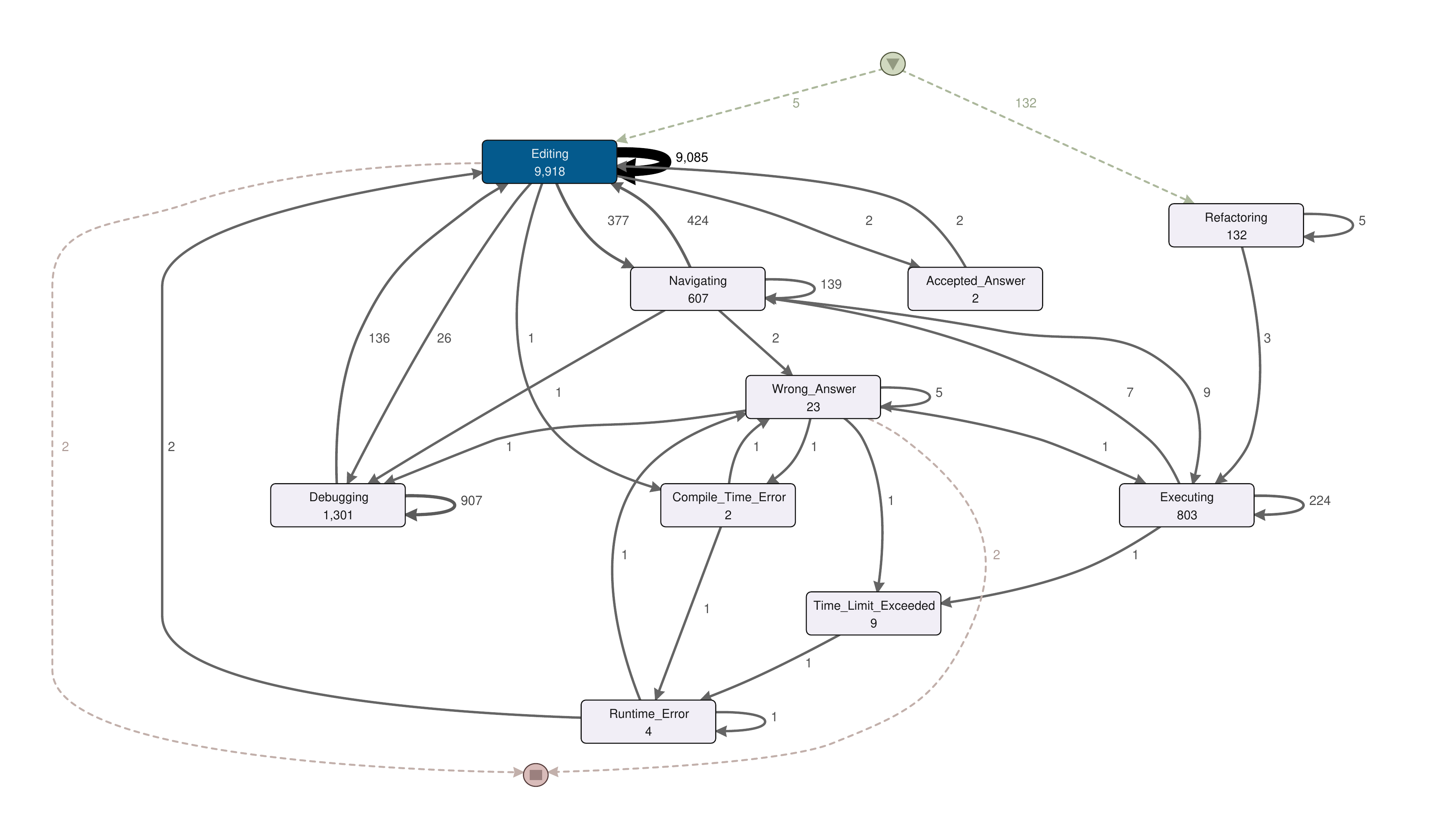}[Process Model characterizing Bottom5 Participants][14][ht!][trim=1cm 1cm 1cm 0cm]












\begin{table}[H]
\tiny
\caption{One-Way ANOVA Results}
\label{table:aov5}
\centering
\begin{tabular}{llccccc}
	\hline\noalign{\smallskip}
	\textbf{} & \textbf{Factor} & \textbf{Df.} & \textbf{Sum Sq.} & \textbf{Mean Sq.} & \textbf{F-value} & \textbf{\textit{p-value}}\\
	\noalign{\smallskip}\hline\noalign{\smallskip}
	\multicolumn{7}{l}{\cellcolor{gray!10}\textbf{Analysis of Variance Test - Top5/Bottom 5/Others}}\\ 
     \textbf{\multirow{1}{*}{\shortstack[c]{Simplicity}}} & \textbf{Performance} & 2 & 0.01150 & 0.005750 &  5.984 & \textbf{0.00594*}\\[0.05cm] 
     & \textbf{Residuals} &  34 & 0.03267 & 0.000961 &  & \\[0.2cm] 
  
     \multicolumn{7}{l}{\cellcolor{gray!10}\textbf{Post Hoc Test}}\\
  
     & \textbf{Treatments} &  \textbf{Diff} & \textbf{Lower} & \textbf{Upper} & \textbf{\textit{p-adj}} & \\[0.05cm]
     & Others-Top5 &  0.01418519 & -0.02279834 & 0.05116871 & 0.6192293  & \\[0.05cm]
     & Bottom5-Top5 & 0.06160000 & 0.01355700 & 0.10964300 & \textbf{0.0094695*}  & \\[0.05cm]
     & Bottom5-Others & 0.04741481 & 0.01043129 & 0.08439834 & \textbf{0.0094774*}  & \\[0.3cm]
     
     \textbf{\multirow{1}{*}{\shortstack[c]{Interactions}}} & \textbf{Performance} & 2 & 1431864 & 715932 &  0.555  & 0.579\\[0.05cm] 
     & \textbf{Residuals} &  34 & 43821404 & 1288865 &  & \\[0.2cm] 
     
     
      \multicolumn{7}{l}{\cellcolor{gray!10}\textbf{Analysis of Variance Test - $>$ Q3(Top 8)/Q1(Bottom 15)/Others}}\\
     \textbf{\multirow{1}{*}{\shortstack[c]{Simplicity}}} & \textbf{Performance} & 2 & 0.00335 & 0.001676 &  1.396 & 0.262\\[0.05cm] 
     & \textbf{Residuals} &  34 & 0.04082 & 0.001201 &  & \\[0.2cm] 
     
      \textbf{\multirow{1}{*}{\shortstack[c]{Interactions}}} & \textbf{Course} & 2 & 165352 & 82676 &  0.062  & 0.94\\[0.05cm] 
     & \textbf{Residuals} &  34 & 45087916 & 1326115 &  & \\[0.2cm] 
 

     \multicolumn{7}{l}{\cellcolor{gray!10}\textbf{Analysis of Variance Test - Top8/Bottom 8/Others}}\\
     \textbf{\multirow{1}{*}{\shortstack[c]{Simplicity}}} & \textbf{Performance} & 2 & 0.00656 & 0.003279 &  2.963 & 0.0651\\[0.05cm] 
     & \textbf{Residuals} &  34 & 0.03762 & 0.001106 &  & \\[0.2cm] 
  
  

     \textbf{\multirow{1}{*}{\shortstack[c]{Interactions}}} & \textbf{Performance} & 2 & 34612 & 17306 &  0.013  & 0.987\\[0.05cm] 
     & \textbf{Residuals} &  34 & 45218656 & 1329960 &  & \\[0.2cm] 
    
    
     \multicolumn{7}{l}{\cellcolor{gray!10}\textbf{Analysis of Variance Test - LEI/ETI/LCD/IGE}}\\
     \textbf{\multirow{1}{*}{\shortstack[c]{Simplicity}}} & \textbf{Course} & 3 & 0.00371 & 0.001237 &  1.009 & 0.401\\[0.05cm] 
     & \textbf{Residuals} &  33 & 0.04046 & 0.001226 &  & \\[0.2cm] 
  
     \textbf{\multirow{1}{*}{\shortstack[c]{Interactions}}} & \textbf{Course} & 3 & 3919333 & 1306444 &  1.043  & 0.386\\[0.05cm] 
     & \textbf{Residuals} &  33 & 41333934 & 1252543 &  & \\[0.2cm]

     \noalign{\smallskip}\hline
   \multicolumn{7}{l}{\textit{*Statistically significant if p-value $<$ 0.05}}\\
   \multicolumn{7}{l}{\shortstack[l]{\textbf{Df.} - Degrees of freedom, \textbf{Sum Sq.} - Sum of Square, \textbf{Mean Sq.} - Mean of Square}}
\end{tabular}
\end{table}



\section{Threats to Validity}
\label{sec:Threatsvalidity5}

The following types of validity issues were considered when interpreting the results of this article. 

\subsection{Construct Validity}
\label{sec:ConstructValidity5}

Construct validity refers to the degree to which inferences can legitimately be made from the operationalizations in a study to the theoretical constructs on which those operationalizations were based.

For operationalizing language models assessment, we used metrics such as perplexity and cross-entropy, and CaoJuan2009, Arun2010, Deveaud2014, and Griffiths2004 to evaluate, from an empirical perspective, the optimal number of topics, and validated their values from multiple perspectives. Other metrics could have been used for the same purpose, such as topic coherence, which may lead to recommending a different optimal number of topics.

Since our sample was not very large, we had to use it for training and test purposes. To strengthen significance, models were trained using 5-fold cross-validation. We are aware that process model metrics such as Precision and Generalization are far from being usable in a more generic process mining context. However, in this study, we were only focused on process simplicity, and regarding that purpose, the mining and tests are valid since we used the same algorithm for all participants.

Each participant used their student number to activate the PyCharm events collector plugin. This approach served as an identification method. 
Events collected and stored in JSON files on developers' devices could have been manually changed. We tried to mitigate this threat of having data tampering by using a hash function on each event at the moment of its creation. As such, each event contains not only information about the IDE activities, but also a hash code introduced as a new property in the event for later comparison with original event data.



\subsection{Internal Validity}
\label{sec:InternalValidity5}
Internal validity refers to the degree of confidence that the causal relationship being tested is trustworthy and not influenced by other factors or variables. 

One typical threat to internal validity related to how subjects are selected. In our case, the population from where our sample was taken, corresponds to all undergraduate students in computer science areas in our university that had attended at least two programming courses. That population was invited to participate by email. The sampling process was the result of a random process of free will where those students that spontaneously decided to participate performed their inscription online. As such, we do not consider this to be a significant validity threat.

Another recurrent internal validity threat is the existence of spurious factors affecting the outcome of the experiment. In mitigation, the programming contest in our study allowed us to block possible confounding factors since they were constant for all subjects: the programming language (Python), IDE (PyCharm), problem complexity (same requirements spec), sprint schedule (4 hours), environment conditions (large shared open space with private tables), and external interference (no contacts were allowed). Once again, we believe that this threat is also not significant.

\subsection{External Validity}
\label{sec:ExternalValidity5}
External validity refers to the extent to which results from a study can be applied (generalized) to other situations, groups or events.

To fully claim that undergraduate students are surrogates of professional programmers, a representative sample of both groups should be assigned the same requirements specification for a Python program, to measure the difference on their outcome. We are not aware of such a study having been published. Nevertheless, there is a likelihood that our students are at least good surrogates for novice professional software developers in Python, because:
\begin{enumerate}[label=(\roman*)]
\item Python has a low learning curve, based on our experience, corroborated by \cite{Nagpal2019Python}, so that the level of proficiency of a professional Python programmer seems to be achievable quickly;
\item our students had attended successfully, on average, two Python courses;
\item a questionnaire filled during inscription showed that participating students, albeit having gone through similar academic paths, had different maturity and skills, as we would expect in professional programmers; that difference most probably will not fade out within the one or two years that will take for the vast majority of these students to become professionals.
\end{enumerate}

\subsection{Conclusion Validity}
\label{sec:ConclusionsValidity5}

The conclusion validity describes our ability to draw statistically correct conclusions based on the measurements. A common threat here is the sample size, but in our case we were able to get a sufficiently large number of subjects to grant statistical significance. 

We carefully evaluated the models perplexity computed to answer RQ2, and the assumption tests to justify the applications of the statistical tests in answering RQ3, however, we have also to accept that our sample is not of large proportions.
We performed an experiment using data from 37 software developers executing well defined and identical programming tasks. Since this is a moderate population size for this type of analysis, we agree this may be a threat to generalize conclusions or make bold assertions. Nevertheless, to our best knowledge, this is the first study involving development sessions and the usage of language models, text and process mining to detect developer's fingerprints during programming tasks. As such, researchers can start from our initial findings and try to falsify our current results and correspondent conclusions.  

\section{Conclusion}
\label{sec:Conclusions5}

\subsection{Main conclusions}

In this work, we tried to understand if fingerprints from development sessions could be extracted from the IDE and an automatic judge platform interactions. Furthermore, we assessed if those sessions could be mined with text mining methods.

We mined the PyCharm and Mooshak events from a group of developers during a Python programming contest aiming to solve six different exercises. Our research regarding development interactions shows that they can be mined as a natural language and using text mining methods with tri- or four-grams being the optimal value for such task. Coherent development fingerprints were discovered and evaluated using process mining methods and correspondent quality metrics. We confirm a significant difference in the process simplicity between the top performers and the ones with unsatisfactory outcomes on the programming exercises.
Results provide evidences to sustain that, to achieve software with good quality, it is not only needed to have developers with the right skills and consistent knowledge about the languages and tools used on their daily tasks. It is also desirable to have developers to follow consistent practices during the development sessions, otherwise, their behaviors may impact the final outcome.





Our approach can be particularly relevant in cases where educators want to assess development profiles within a group of students, before and after classes are given. It can be also a valid approach to measure and monitor productivity within and between software teams. As we showed, by analyzing the development sessions fingerprints and complexity, non efficient developers can easily be detected.

Last generation IDEs provide a plethora of functionalities, such as code completion, automated packaging and optimized continuous integration features to assist programmers on their daily activities. However, these IDEs do not guide developers on their coding practices. The fingerprints we detected, either classified as good or bad practices\footnote{We already called them - software development process smells}, may be used as a trigger for IDE vendors to evaluate the possibility to include additional intelligence in their tools, such as task/workflow monitoring and suggesting program runs, identify testing slots and appropriately recommend debugging and refactoring actions along a development session.

\subsection{Future work}
\label{sec:FutureWork}

\subsubsection{Automation}
We are still scratching the surface in mining developers' activities. Existing mining tools are not ready yet to automate the complete flow of: collect and pre-process data, discover processes behaviors, compute metrics, and export results. Further work is required to set up a pipeline capable of providing just-in-time feedback, both to software developers, to provide self-awareness on performance/behavior, as to software project managers, since the profile of team members allows a more informed resource allocation.

\subsubsection{Low and No Code paradigms} 
Novel software development paradigms, such as low and no code, shift the focus from the textual programming and put it into the visual artifacts and components from which modern applications are built upon. 
Our work fits well in cases where textual programming is banned, giving rise to the so-called \textit{citizen developers}, and therefore, most likely to distinct development processes and coding behaviors when compared with conventional programming practices.
We plan to perform additional experiments using low or no code platforms and assess developers' process fingerprints and overall behaviors.








\section*{Acknowledgement}
This work was partially funded by the Portuguese Foundation for Science and Technology, under ISTAR-Iscte projects UIDB/04466/2020 and UIDP/04466/2020, and INESC-ID project UIDB/50021/2020.




\bibliography{references,references-offline}



\end{document}